%% file: v3_Adaptive.tex
\documentclass[12pt,onecolumn,letterpaper]{IEEEtran}
\input{preamble}

\usepackage{graphicx,cite}
\usepackage{enumerate}
\usepackage{color}
\usepackage[T1]{fontenc}
\usepackage{aecompl}
\usepackage{booktabs}
\usepackage{tabularx}
\usepackage{dsfont}
\usepackage{lettrine}
\usepackage[margin=1in]{geometry}
\IEEEoverridecommandlockouts

\makeatletter

\newcommand{\Rmnum}[1]{\expandafter\@slowromancap\romannumeral #1@}
\makeatother

\begin{document}
\title{Adaptive Beam Alignment using Noisy Twenty Questions Estimation with Trained Questioner}

\author{
\IEEEauthorblockN{Chunsong Sun and Lin Zhou} \\
\IEEEauthorblockA{
Beihang University\\
Emails: \{sunchunsong, lzhou\}@buaa.edu.cn} \\
\thanks{}
}

\maketitle

\begin{abstract}
The 6G communication systems use millimeter-wave and multiple-input multiple-output technologies to achieve wide bandwidth and high throughput, leading to an indispensable need for beam alignment to overcome severe signal attenuation. Traditional sector-search-based beam alignment algorithms rely on sequential sampling to identify the best sector, resulting in a significant latency burden on 6G communication systems. Recently proposed adaptive beam alignment algorithms based on the active learning framework address the problem, aiming to identify the optimal sector with the fewest possible samples under the same sector partition. Nevertheless, these algorithms either lack feasibility (Chiu, Ronquillo and Javidi, JSAC 2019) due to ideal assumptions or lack interpretability (Sohrabi, Chen and Yu, JSAC 2021) due to the use of end-to-end black-box neural networks. To avoid ideal assumptions and maintain interpretability, in this paper, we address all above problems by proposing an adaptive beam alignment algorithm using the framework of noisy twenty questions estimation with a trained questioner. Specifically, we use two methods for training the questioner to eliminate reliance on ideal assumptions. The first method maps queries of twenty questions estimation to beamforming vectors via weighted summation of steering vectors, as an initial attempt to address the feasibility problem encountered in prior pioneering study by Chiu, Ronquillo and Javidi (JSAC 2019). The second method uses multi-layer fully connected neural networks to achieve improved performance while only employing them to train the questioner, which can effectively mitigate the interpretability issues in prior study by Sohrabi, Chen and Yu (JSAC 2021). Furthermore, we provide numerical simulations to illustrate the effectiveness of our proposed adaptive beam alignment algorithms and demonstrate that our algorithms outperform all benchmark algorithms.

\end{abstract}


\section{Introduction}
\label{sec_intro}

The sixth generation (6G) mobile communication is critical infrastructure to meet the ever-growing demand for wireless data traffic in future communication systems~\cite{tataria20216g}. Massive multiple-input multiple-output (MIMO) and millimeter-wave (mmWave) are two technical pillars that support 6G communication systems. High-frequency mmWave provides abundant spectrum resources and wide communication bandwidth~\cite{wang2018millimeter}. Because of short wavelengths, mmWave communication equipment is highly portable due to small antenna size. However, mmWave communication can be easily affected by environmental factors such as atmosphere, precipitation, and suspended solids in the air~\cite{nagaraj2018impact}, leading to higher transmission loss compared to lower frequency communications. Furthermore, mmWave has poor diffraction capability, which further limits its propagation distance~\cite{niu2015survey}. To address these problems, beamforming for MIMO communications can be used to generate directional beams to extend communication distance and improve communication quality. A critical initial step to achieve beamforming is beam alignment (BA), where one identifies the optimal beam direction to ensure best communication performance.

Traditional sector-search-based BA algorithms include naive exhaustive search and hierarchical exhaustive search with a fixed codebook of given beam directions. In the naive exhaustive BA algorithm, the transmitter (TX) periodically sends pilot signals, and the receiver (RX) selects the optimal beam direction based on the received pilot signals using criteria such as the maximum received signal strength indicator. However, this algorithm needs to sample each sector sequentially, which results in a significant latency burden. Hierarchical exhaustive BA algorithms, such as those specified by IEEE 802.11 ad~\cite{80211}, IEEE 802.15.3c~\cite{80215} and 5G NR~\cite{ZTE2017OnCSI}, partially address the above problem by progressively reducing, over multiple stages, the number of sectors that need to be searched. Nevertheless, since the codebook is fixed rather than adaptively generated, the hierarchical approach provides only limited performance improvement and suffers from error propagation.

To minimize the number of samples, Chiu, Ronquillo and Javidi~\cite{chiu2019active} adopted the idea of active learning, specifically noisy twenty questions estimation pioneered by R\'enyi~\cite{renyi1961problem} and Ulam~\cite{ulam1991adventures}. Twenty questions estimation is a question and answer style game between two players: one player called the oracle picks something as a secret and the other player called the questioner aims to accurately guess the secret by posing as fewer queries as possible. In the BA algorithm of~\cite{chiu2019active}, at each time point, designing a beamforming vector corresponds to designing a query in twenty questions estimation and one measurement result of the received signal corresponds to a response from an oracle in twenty questions. Two measurement rules are considered in~\cite{chiu2019active}: full measurement and 1-bit measurement. With full measurement, one uses the complete information of received signals to estimate the angle-of-arrival (AoA), while with 1-bit measurement, only extreme quantized information of received signals can be obtained. Using the connection between BA and twenty questions estimation, Chiu \emph{et al.} used a hierarchical posterior matching strategy to generate queries, converted queries into beamforming vectors with a pre-designed ideal beamforming codebook, and estimated the AoA using noisy responses and decoding function of twenty questions. Subsequently, many studies applied the idea of active learning to BA via other tools of machine learning, such as neural network (NN)~\cite{sohrabi2021deep,jiang2023active,cheng2024deep} and multi-armed bandits (MAB)~\cite{wei2022fast,zhang2020beam,khirwadkar2023max,blinn2025mab,blinn2021mmwave}. In particular, the authors integrated the main steps of BA, including the design of beamforming vectors, the update of posterior probability vectors and the estimation of AoA, into a deep neural network (DNN) for training. The authors of~\cite{jiang2023active} focused on the design of beamforming vector using a recurrent neural network (RNN) with long short-term memory. In~\cite{wei2022fast}, the authors proposed a hierarchical exhaustive algorithm using multi-armed bandits by modeling each narrow beam direction as a base arm, combining several adjacent base arms as a super arm with wide beam direction and proposing a two-phase track-and-stop algorithm that firstly pulls super arms and then pulls base arms.

However, all the above algorithms have limitations. Specifically, the BA algorithm in~\cite{chiu2019active} did not specify how the ideal beamforming vector in~\cite[Assumption 4]{chiu2019active} is designed, making it difficult to apply in practical communication systems. Although the studies in~\cite{sohrabi2021deep,jiang2023active} addressed the above feasibility problem, the extensive use of DNNs or RNNs reduces the interpretability of the algorithms, in the sense that their performance can neither be explained by existing theoretical results in~\cite{chiu2021low} nor supported by new theoretical analyses. Although the algorithms based on MAB~\cite{wei2022fast,zhang2020beam} had good interpretability, they require frequent arm pulling, which implies frequent sector searches, thereby leading to undesired high latency overhead. In summary, existing algorithms suffer from issues of feasibility, interpretability or high latency. One might wonder whether it is possible to propose a feasible and interpretable BA algorithm that has low latency and good performance. In this paper, we answer the question affirmatively. Our main contributions are summarized as follows.

\subsection{Main Contributions}
\label{intro Main Contributions}

To simultaneously achieve feasibility, interpretability and low latency, we propose adaptive BA algorithms using noisy twenty questions estimation with two methods to train the questioner and generate beamforming vectors.

\begin{itemize}

\item To address the infeasibility problem of the BA algorithm in~\cite{chiu2019active} due to the assumption of perfect beamforming vectors (cf.~\cite[Assumption 4]{chiu2019active}), we abandon this assumption and adopt a linear weighted sum (LWS) method to convert queries into practical beamforming vectors. In particular, for each query, we select a number of beam directions with equal intervals within the search region, calculate the weighted summation of the steering vectors corresponding to the angles of beam directions, and use the summation as the beamforming vector. With the proposed LWS-based method, our adaptive BA algorithm can be directly applied to practical communication systems. In addition, it helps us gain preliminary insights into why performance degrades and to what extent when the ideal assumption is not met.

\item Inspired by~\cite{sohrabi2021deep}, we employ a DNN to facilitate the transformation from queries to beamforming vectors, with the goal of achieving less performance degradation compared with the LWS-based method. Specifically, we calculate the steering vectors of angles within the search region and feed them as inputs into a multi-layer fully connected NN. Compared with the LWS-based method, the DNN-based method achieves better performance in numerical simulations. Furthermore, compared with the BA algorithm in~\cite{sohrabi2021deep} which adopts an end-to-end DNN architecture, our algorithm improves interpretability by limiting the role of the DNN in the active-learning-based BA algorithm.

\item We next investigate the following questions via numerical simulations of the two proposed methods in Figs. \ref{lws_vs_dnn_1bit}-\ref{full_exp}: what limits active-learning-based beam alignment algorithms from achieving their theoretical optimum; why the questioner trained by the DNN-based method and the full measurement rule can exhibit better performance; and whether the size of the search region affects the performance of the trained questioner.

\item Finally, we run comprehensive numerical examples to demonstrate the performance of our BA algorithms in Fig. \ref{quadratic_loss}. In our numerical simulations, we first show that the performance of our BA algorithm under the 1-bit measurement rule is better than the naive beam sweeping algorithm and hierarchical beam sweeping algorithm~\cite{ZTE2017OnCSI} under the full measurement rule. Furthermore, under the full measurement rule, we run extensive numerical examples to compare the performance of our algorithms, the naive beam sweeping algorithm, hierarchical beam sweeping algorithm~\cite{ZTE2017OnCSI} and the end-to-end DNN-based BA algorithm~\cite{sohrabi2021deep}. Simulation results show that i) the performance of our BA algorithms under the full measurement rule is much better than under the 1-bit measurement rule and ii) under the full measurement rule, our proposed BA algorithms outperform all benchmark BA algorithms.

\end{itemize}

\subsection{Organization of the Rest of the Paper}

The rest of this paper is organized as follows. Section \ref{sec_pf} introduces the system model for adaptive BA, describes the noisy twenty questions estimation framework, and introduces the BA problem using twenty questions for practical communication systems. Section \ref{sec_mr} presents our two proposed methods to train questioners to generate feasible beamforming vector based on queries of twenty questions, and numerically analyze the performance of the two proposed methods. Section \ref{sec_nr} provides extensive numerical simulation results to verify the validity of query-dependent channel models for the adaptive BA algorithms and to demonstrate the superior performance of our proposed BA algorithms over existing benchmark algorithms. Finally, Section \ref{sec_conc} concludes the paper and discusses future directions.

\subsection*{Notation}

Random variables and their realizations are denoted by upper case variables (e.g., $X$) and lower case variables (e.g., $x$), respectively. All sets are denoted in calligraphic font (e.g., $\calX$). We use $\bbC, \bbR, \bbR_+$ and $\bbN$ to denote the set of complex numbers, real numbers, positive real numbers, and positive integers, respectively. For any closed interval $\calI\subset \bbR$, we use $\bbP(\calI)$ to denote its power set that includes all possible subsets of $\calI$. Given any three positive integers $(l, m, n)\in\bbN^3$, we use $[m:l:n]$ to denote the set $\{m, m + l,\ldots,m+\lfloor \frac{n-m}{l} \rfloor l\}$, use $[m : n]$ to denote $[m:1:n]$, and use $[m]$ to denote $[1 : m]$. Fix an integer $m\in\bbN$. For any length-$m$ vector $\ba:=(a_1, \ldots , a_m)$, the infinity norm is defined as $\|\ba\|_\infty := \max_{i \in [m]}|a_i|$, and the Euclidean norm is defined as $\|\ba\|_2 := \sqrt{\sum_{i \in [m]} a_i^2}$. We use $\{\mathbf{n}\}_m$ to denote the all-$n$ vector of length $m$ for any $n\in\bbR_+$ and use $\mathbf{I}_m$ to denote the identity matrix with dimensions $m$. Given two alphabets $(\calX,\calY)$, we use $\calF(\calX)$ to denote the set of all probability distributions on the set $\calX$ and use $\calP(\calY|\calX)$ to denote the set of all conditional probability distributions from $\calX$ to $\calY$. Given any $(p,q)\in(0,1)\times\{0,1\}$, let Bern$(p)$ denote the Bernoulli distribution with parameter $p$, and Bern$(q,p)$ denote the 0-1 distribution with parameter $(p,q)$, i.e., Bern$(q;p):= p^q(1-p)^{1-q}$. We use $\mathcal{CN}(\mu,\Sigma)$ to denote the multivariate complex Gaussian distribution with mean vector $\mu$ and covariance matrix $\Sigma$, and use $\mathcal{CN}(\cdot,\mu,\Sigma)$ to denote its probability density function. Finally, we use $\mathds{1}(\cdot)$ to denote the indicator function.

\section{Beam Alignment via Twenty Questions Estimation} 
\label{sec_pf}

This section describes the beam alignment algorithm in \cite{chiu2019active} based on twenty questions estimation. Specifically, we first describe the system model for adaptive BA in Section \ref{pf System Model for Adaptive Beam Alignment}, then introduce noisy twenty questions estimation in Section \ref{pf Noisy Twenty Questions Estimation Framework} and finally clarify how twenty questions estimation can be used in the beam alignment algorithm in Section \ref{pf Adaptive Beam Alignment via Noisy Twenty Questions Estimation Framework}.

\subsection{Adaptive Beam Alignment}
\label{pf System Model for Adaptive Beam Alignment}

\begin{figure}[tb]
\centering
\includegraphics[width=0.6\columnwidth]{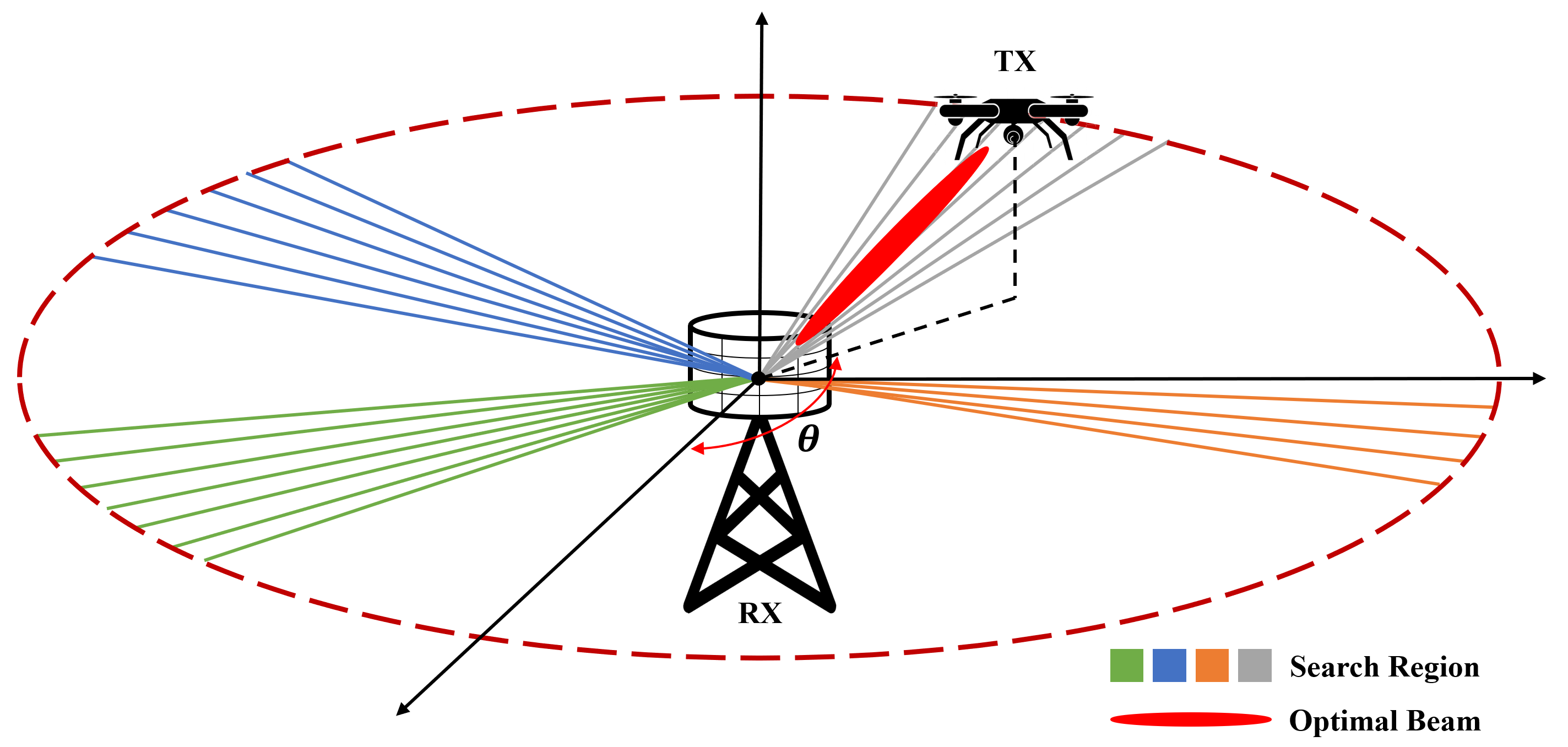}
\caption{Illustration of an adaptive beam alignment procedure in an air-ground communication scenario with a BS and a UAV.}
\label{Theta}
\end{figure}

This subsection reviews the concepts of adaptive BA from~\cite{chiu2019active,sohrabi2021deep,sun2025resolutionlimitsnonadaptive20}. Consider an air-ground communication scenario with a base station (BS) that serves as a RX and a unmanned aerial vehicles (UAV) that serves as a TX, illustrated in Fig. \ref{Theta}. Assume that RX is equipped with a uniform linear array with $N_R \in \bbN$ antennas and TX is a single-antenna UAV. More generally, there are two approaches that are applicable when the transmitter is also equipped with multiple antennas. One approach is for the receiver to perform omnidirectional reception to identify the optimal sector for the transmitter. Then the roles are swapped, and this process is repeated until the sector is narrowed down to the required precision. The other approach involves iteratively searching for the beamforming vectors of both the transmitter and the receiver, inspired by~\cite{jiang2023active}. We do not discuss this in detail here, as it is not the main focus of this paper. At any time slot $t\in \bbN$, to establish a reliable communication link between TX and RX by BA, TX sends a pilot $x_t \in \bbC$ with a power constraint $P \in \bbR_+$.

Let $[\theta_{\mathrm{min}},\theta_{\mathrm{max}}] \subseteq [-180^{\circ},180^{\circ}]$ be the angle range where the pilot signal from TX may be located. Given the antenna spacing $d \in \bbR_+$ and the wavelength $\lambda \in \bbR_+$, the steering vector with the AoA in azimuth $\theta \in [\theta_{\mathrm{min}},\theta_{\mathrm{max}}]$ can be described as:
\begin{align}
\bA(\theta)&:=\big(A_1(\theta),\ldots,A_{N_R}(\theta)\big) =\sqrt{\frac{1}{N_R}}\Big(1,e^{j\frac{2 \pi d}{\lambda}\sin\theta},\ldots,e^{j(N_R-1)\frac{2 \pi d}{\lambda}\sin\theta}\Big). \label{steering_vector}
\end{align}
Given the fading coefficient $\alpha \in \bbC$, the small-scale channel can be described as:
\begin{align}
\bh:=\alpha \bA(\theta).
\end{align}
Consistent with~\cite{ronquillo2023integrated}, we assume that an air-ground communication scenario is dominated by the line-of-sight (LoS) path without obstructions and reflectors. At the time point $t$, if the beamforming vector set of RX is $\bw_t \in \bbC^{N_R}$, the received signal is given by
\begin{align}
\by_t &:= \sqrt{P}\bw_t^H \bh x_t + \bw_t^H \bn_t = \alpha\sqrt{P}\bw_t^H \bA(\theta) x_t + \bw_t^H \bn_t , \label{r-signal}
\end{align}
where $\bn_t \sim \mathcal{CN}(\{\mathbf{0}\}_{N_R},\sigma^2\mathbf{I}_{N_R})$ is the independent Gaussian noise vector with homogeneous variance $\sigma^2 \in \bbR_+$. To facilitate the subsequent analysis, we temporarily assume $\alpha = P = 1$. In Section \ref{nr The Case of Unknown Fading Coefficient}, we will address the case of unknown $\alpha$ using a Kalman filter estimator.

\begin{figure}[tb]
\centering
\includegraphics[width=0.4\columnwidth]{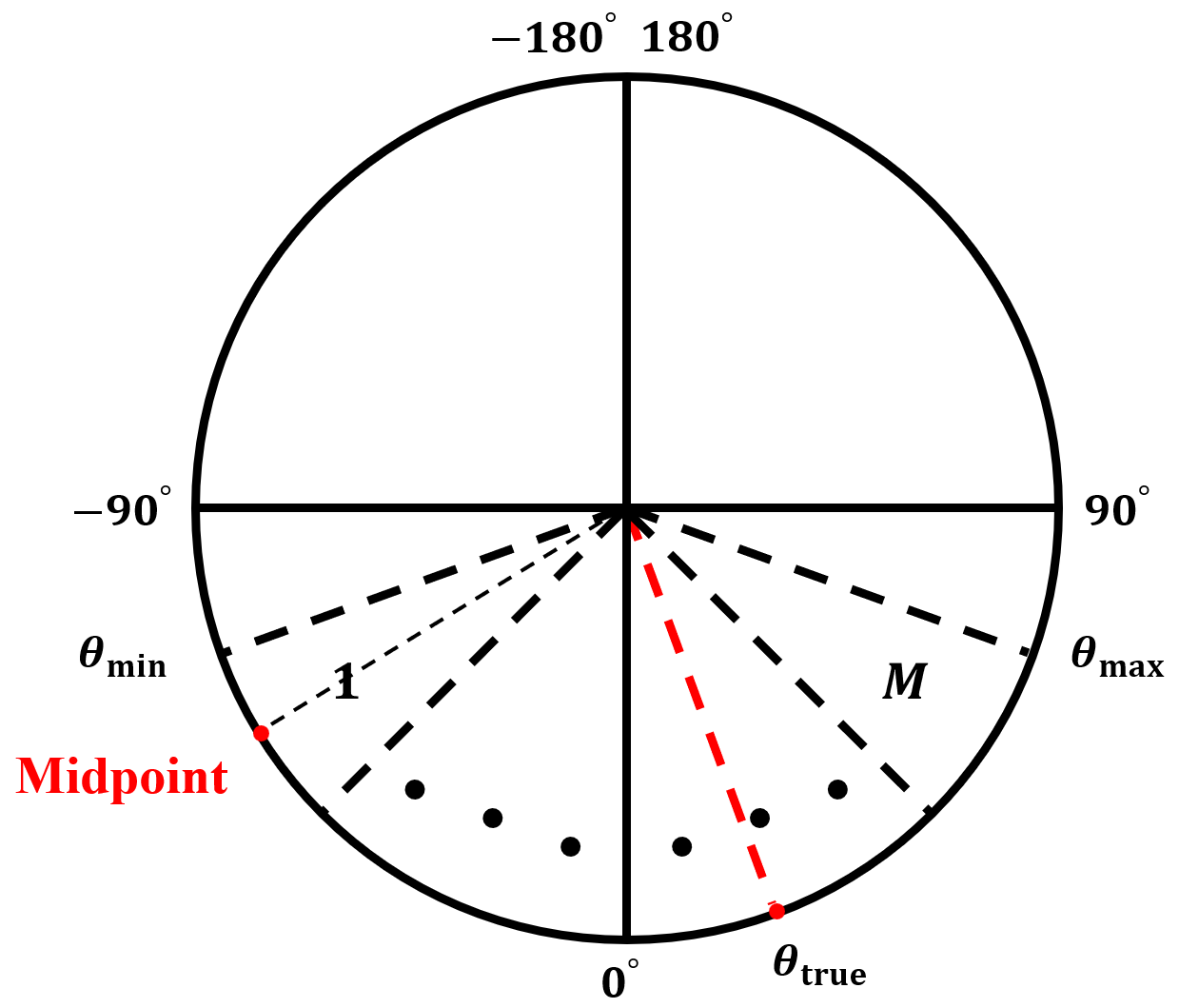}
\caption{Illustration of sub-intervals of search interval $[\theta_{\mathrm{min}},\theta_{\mathrm{max}}]$.}
\label{subinterval_angle}
\end{figure}

Assume that TX remains stationary for a period of time and sends pilot signals. In BA, the goal of RX is to accurately estimate the AoA of signals from TX using as few pilot signals as possible. The crux is to adaptively adjust the beamforming vector until the AoA of pilot signals can be estimated accurately. Specifically, we can divide the whole search interval $[\theta_{\mathrm{min}},\theta_{\mathrm{max}}]$ into $M \in \bbN$ sub-intervals, where the true AoA $\theta_{\mathrm{true}} \in [\theta_{\mathrm{min}},\theta_{\mathrm{max}}]$ is located within one of these sub-intervals, as shown in Fig. \ref{subinterval_angle}. Because of the limitation imposed by the number of array antennas $N_R$, the resolution of the adaptive BA $\frac{1}{M}$ cannot be arbitrarily small, so we set $M \leq N_R$. At time point $t$, for any $i \in [M]$, let $\rho_{t,i}$ denote the the probability that the AoA lies in the $i$-th sub-interval, and define the probability vector $\bm{\rho}_t:=(\rho_{t,1},\ldots,\rho_{t,M})$.

Using the probability vector $\bm{\rho}_t$, RX generates an angular search region $\calA_t^{\mathrm{angle}} \subseteq [\theta_{\mathrm{min}},\theta_{\mathrm{max}}]$ using a mapping $\Gamma_\rma(\cdot):~(0,1)^M \rightarrow \bbP([\theta_{\mathrm{min}},\theta_{\mathrm{max}}])$. Subsequently, RX generates the corresponding beamforming vector $\bw_t$ with $\calA_t^{\mathrm{angle}}$ using another mapping $\Gamma_\rmb(\cdot):~\bbP([\theta_{\mathrm{min}},\theta_{\mathrm{max}}]) \rightarrow \bbC^{N_R}$. Let $\Gamma_\rmb^{-1}(\cdot)$ denotes the inverse mapping of $\Gamma_\rmb(\cdot)$. The above two mapping functions $\Gamma_\rma(\cdot)$ and $\Gamma_\rmb(\cdot)$ are critical for adaptive BA. Constructing feasible mapping functions is the focus and main contributions of this paper.

Applying the beamforming vector $\bw_t$, RX processes the received signal $\by_t$ with one of the following two measurement rules~\cite[Section III-D]{chiu2019active}:
\begin{enumerate}
\item full measurement rule
\begin{align}
\Gamma_\rmm(\by_t):=\by_t. \label{full}
\end{align}
\item 1-bit measurement rule
\begin{align}
\Gamma_\rmm(\by_t):=\mathds{1}\big(|\by_t|^2 > \iota\big), \label{1bit}
\end{align}
where $\iota \in \bbR_+$ denotes the power threshold. 
\end{enumerate}
The full measurement rule, serving as a representative of ideal communication conditions, leads to better performance as the received signal can be fully used, while the 1-bit measurement rule simplifies the theoretical analysis and represents one of the most extreme communication conditions.

To update the posterior probability vector $\bm{\rho}_t$ with the measurement result $\Gamma_\rmm(\by_t)$, we first define the following function $\Delta(\cdot)$ of the received signal $\by_t$:
\begin{align}
\Delta(\Gamma_\rmm(\by_t),\theta_{\mathrm{true}},\theta_i,\bw_t)
&:=\left\{
\begin{aligned}
&\mathcal{CN}\big(\Gamma_\rmm(\by_t);\alpha \sqrt{P} \bw_t^HA(\theta_i),\sigma^2\big),&\text{Full Measurement Rule}, \\
&\mathrm{Bern}\big(z_t \oplus \mathds{1}\big(\theta_i \in \Gamma_\rmb^{-1}(\bw_t)\big);p\big) ,&\text{1-bit Measurement Rule},
\end{aligned}
\right. \label{Delta}
\end{align}
where
\begin{align}
z_t := \mathds{1}\big(\theta_{\mathrm{true}} \in \Gamma_\rmb^{-1}(\bw_t)\big) \oplus u_t(p) \label{z_t}
\end{align}
denotes the noisy measurement result indicating whether the true AoA $\theta_{\mathrm{true}}$ lies within the angular search region $\calA_t^{\mathrm{angle}}$, with the equivalent Bernoulli noise $u_t(p) \sim \mathrm{Bern}(p)$ with flipping probability $p$ and $\oplus$ being the exclusive OR operator. Using function $\Delta(\cdot)$ in \eqref{Delta}, RX updates the posterior probability vector $\bm{\rho}_{t+1}$ as follows:
\begin{align}
\rho_{t+1,i}:=\frac{\rho_{t,i}\Delta(\Gamma_\rmm(\by_t),\theta_{\mathrm{true}},\theta_i,\bw_t)}{\sum_{i^{\prime} \in [M]}\rho_{t,i^{\prime}}\Delta(\Gamma_\rmm(\by_t),\theta_{\mathrm{true}},\theta_{i^{\prime}},\bw_t)}, \forall~i \in [M], \label{update}
\end{align}
where $\theta_i$ denotes the angle corresponding to the midpoint of the $i$-th sub-interval.

Fix a tolerable error probability $\varepsilon \in (0,1)$ and maximum number of allowable iterations $n \in \bbN$. If $n$ is reached or if the posterior probability of a certain sub-interval is greater than $1-\varepsilon$ at a time point, the BA algorithm stops and the RX claims the $\hat{i}_\tau$-th sub-interval as the sub-interval where the AoA is located, and chooses the center of the sub-interval as the estimate of AoA, where
\begin{align}
\hat{i}_\tau:=\argmax_{i \in [M]} \rho_{\tau,i},
\end{align}

\begin{figure}[tb]
\centering
\includegraphics[width=0.6\columnwidth]{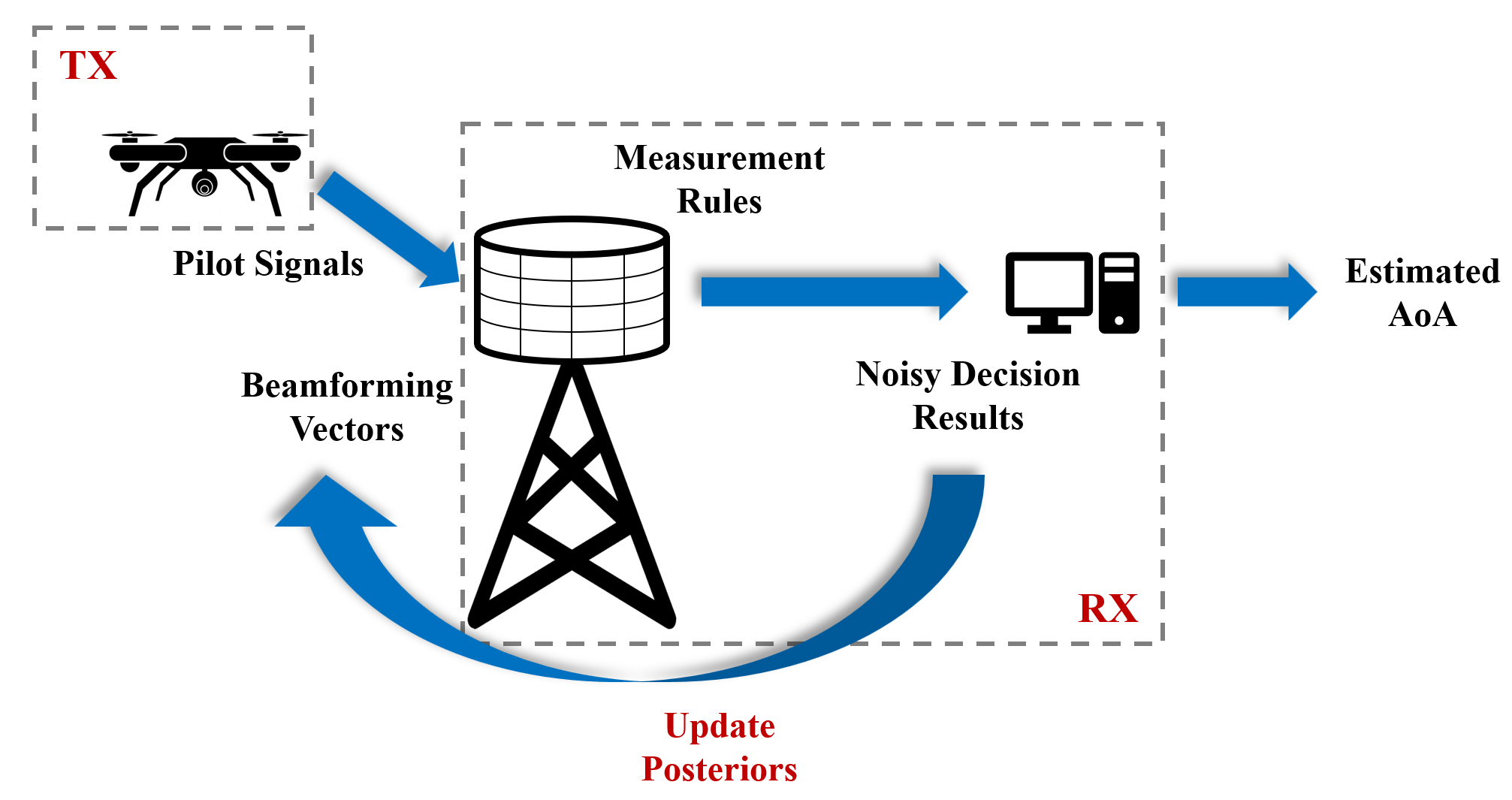}
\caption{Diagram of adaptive BA algorithm.}
\label{adaptiveBA}
\end{figure}

The above adaptive BA algorithm (cf. Fig. \ref{adaptiveBA}) can be modeled as a random variable estimation problem, where the AoA $\theta_{\mathrm{true}}$ is an arbitrary random variable to be estimated within a certain interval, the beamforming vector $\bw_t$ at each time point $t$ corresponds to a query used to check whether the random variable lies in a certain range, and the received signal $\by_t$ corresponds to a noisy response to the query. Therefore, noisy twenty questions estimation framework~\cite{zhou2021achievable,sun2023achievable,zhou2025twenty} can be used in the BA algorithm.

\subsection{Noisy Adaptive Twenty Questions Estimation} 
\label{pf Noisy Twenty Questions Estimation Framework}

\begin{figure}[tb]
\centering
\includegraphics[width=0.8\columnwidth]{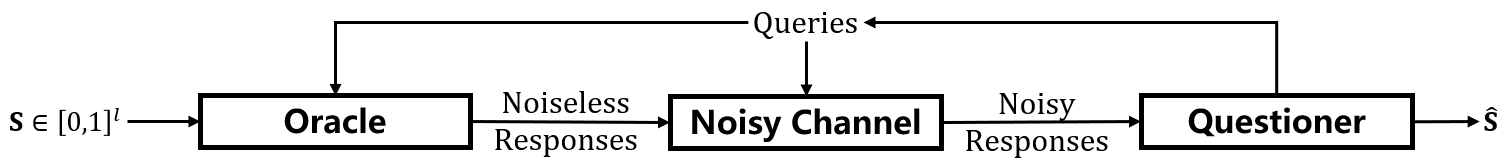}
\caption{Diagram of noisy twenty questions estimation framework}
\label{20Qs_framework}
\end{figure}

This subsection introduces the noisy twenty questions estimation framework~\cite{zhou2025twenty}. Let $\bS=(S_1,\ldots,S_l) \in [0,1]^l$ be an $l$-dimensional random variable generated from an arbitrary distribution $f_{\bS} \in \calF(\calS)$. As shown in Fig. \ref{20Qs_framework}, a questioner aims to estimate a random variable $\bS$ with the assistance of an oracle who has access to $\bS$. At each time point $t \in \bbN$, the questioner generates a Lebesgue measurable query $\calA_t \subseteq [0,1]^l$ and asks the oracle whether $\bS$ lies in the set $\calA_t$. The query set $\calA_t$ is generated using all previous queries $(\calA_1,\ldots,\calA_{t-1})$ and their noisy responses $(Y_1,\ldots,Y_{t-1})$. After receiving $\calA_t$, the oracle generates binary yes/no answers $X_t=\mathds{1}(\bS \in \calA_t)$ and passes $X_t$ over a query-dependent channel~\cite[Eq. (1)]{zhou2021resolution} with transition probability $P_{Y_t|X_t}^{\beta(|\calA_t|)} \in \calP(\calY|\calX)$, yielding a noisy response $Y_t$, where $\beta(\cdot)$ is a bounded Lipschitz continuous function, and $|\calA_t|$ denotes the size of the query $\calA_t$. For example, a query-dependent binary symmetric channel (BSC) is defined as follows.

\begin{definition} \label{def_bsc}
Given any $\calA \subseteq [0, 1]$, a channel with crossover probability $P_{Y|X}^{\beta(|\calA|)}$ is said to be a query-dependent BSC if $\calX = \calY = \{0, 1\}$ and $\forall~ (x,y) \in \{0, 1\}^2$,
\begin{align}
P_{Y|X}^{\beta(|\calA|)} := \big(\beta(|\calA|)\big)^{\mathds{1}(y \neq x)}\big(1-\beta(|\calA|)\big)^{\mathds{1}(y = x)}. \label{bsc}
\end{align} 
\end{definition}
Under the query-dependent BSC, the query size function $\beta(\cdot)$ determines its crossover probability. Smaller size query sets correspond to lower crossover probability when $\beta(\cdot)$ is an increasing function. The query-dependent BSC can be used to model the estimation accuracy of the BA algorithm under the 1-bit measurement rule.

And a query-dependent additive white Gaussian noise (AWGN) channel defined as follows.
\begin{definition} \label{def_awgn}
Given any $\calA \subseteq [0, 1]$, a channel with transition probability $P_{Y|X}^{\beta(|\calA|)}$ is said to be a query-dependent AWGN channel with parameter $\sigma\in\bbR_+$ if $\calX = \calY = \bbR$ and $\forall~ (x,y) \in \bbR^2$,
\begin{align}
P_{Y|X}^{\beta(|\calA|)} := \frac{1}{\sqrt{2\pi(\beta(|\calA|)\sigma)^2}}\exp\Big(-\frac{(y-x)^2}{2(\beta(|\calA|)\sigma)^2}\Big). \label{awgn}
\end{align}
\end{definition}
Under the query-dependent AWGN channel, the query size function $\beta(\cdot)$ corresponds to noise variance. Smaller query sets $\calA$ correspond to higher signal to noise ratio when $\beta(\cdot)$ is an increasing function. The query-dependent AWGN can be used to model the estimation accuracy of the BA algorithm under the full measurement rule.

With responses to all queries $(Y_1,\ldots,Y_t)$, the questioner determines whether to stop the query procedure. If yes, the questioner generates an estimate $\hat{\bS}$. Otherwise, the query procedure continues. A noisy adaptive query procedure can be defined as~\cite[Definition 2]{zhou2021achievable}.

\subsection{Adaptive Beam Alignment via Noisy Adaptive Twenty Questions Estimation}
\label{pf Adaptive Beam Alignment via Noisy Twenty Questions Estimation Framework}

\begin{figure}[tb]
\centering
\includegraphics[width=0.8\columnwidth]{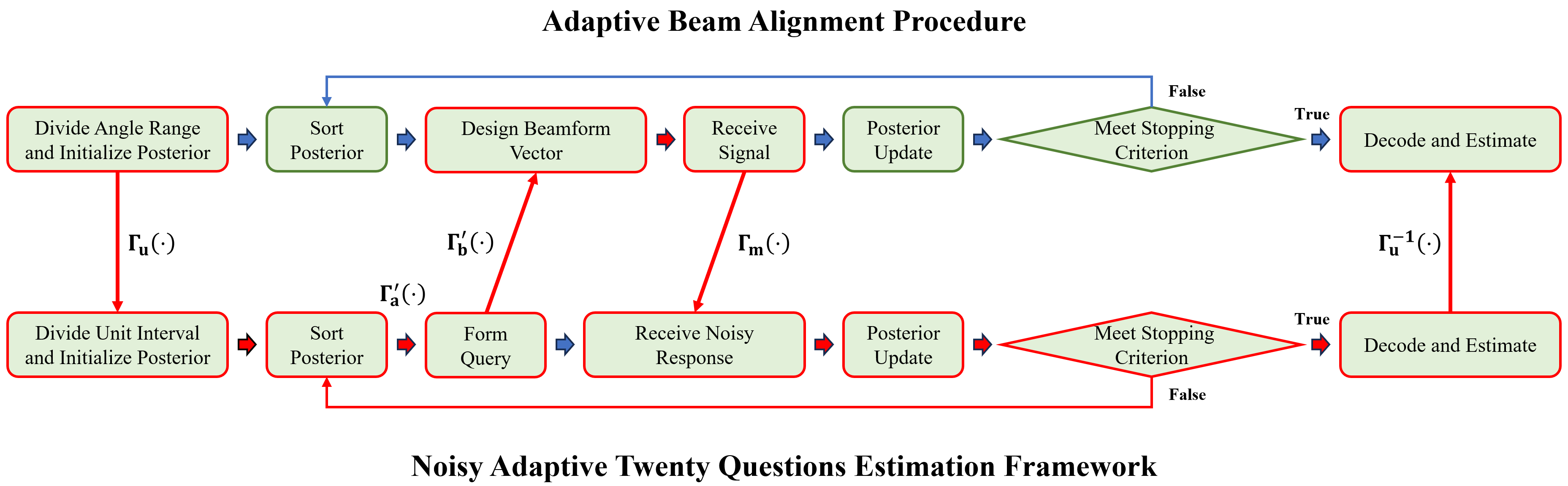}
\caption{Diagram of conversion relationship between adaptive BA algorithm and noisy adaptive twenty questions estimation framework}
\label{Conversion}
\end{figure}

In the end of Section \ref{pf System Model for Adaptive Beam Alignment}, we show that the adaptive BA problem is analogous to adaptive twenty questions estimation (cf. Fig. \ref{adaptiveBA}). This section details the steps for constructing a BA algorithm using a query procedure for twenty questions, as illustrated in Fig. \ref{Conversion}, where $\Gamma_\rma^{\prime}(\cdot):~(0,1)^M \rightarrow \bbP([0,1])$ is a mapping function to convert the posterior probability vector into a query set, and $\Gamma_\rmb^{\prime}(\cdot):~\bbP([0,1]) \rightarrow \bbC^{N_R}$ is a mapping function to convert a query set into a beamforming vector.

The first step is to map the angular range $[\theta_{\mathrm{min}},\theta_{\mathrm{max}}]$ of RX into the unit interval $[0,1]$. For any AoA $\theta_{\mathrm{true}} \in [\theta_{\mathrm{min}},\theta_{\mathrm{max}}]$, the mapping function $\Gamma_\rmu(\cdot)$ defined as
\begin{align}
\Gamma_\rmu(\theta_{\mathrm{true}}) := \frac{\theta_{\mathrm{true}}-\theta_{\mathrm{min}}}{\theta_{\mathrm{max}}-\theta_{\mathrm{min}}},
\end{align}
where $\Gamma_\rmu(\cdot)^{-1}$ denotes its inverse mapping function.
Next we partition the unit interval into $M$ equal-sized non-overlapping sub-intervals $(\calC_1,\ldots,\calC_M)$, and define the quantization function $\Gamma_\rmq(\cdot)$ as
\begin{align}
\Gamma_\rmq(\Gamma_\rmu(\theta_{\mathrm{true}})) := \lceil \Gamma_\rmu(\theta_{\mathrm{true}}) \times M \rceil.
\end{align}

Then we follow the sortPM strategy proposed in~\cite[Section III-A]{chiu2021low} and do not elaborate further due to space limitations. Once we obtain an estimate of the sub-interval in which the target lies, we can map it to $\hat{\theta}$ as the estimate of AoA with $\Gamma_{\rmu}^{-1}(\cdot)$. To measure the accuracy of the estimation, we define the quadratic loss $\delta^2$ as follows:
\begin{align}
\delta^2 := \big(\hat{\theta}-\theta_{\mathrm{true}}\big)^2.
\end{align}

\begin{figure}[tb]
\centering
\includegraphics[width=0.8\columnwidth]{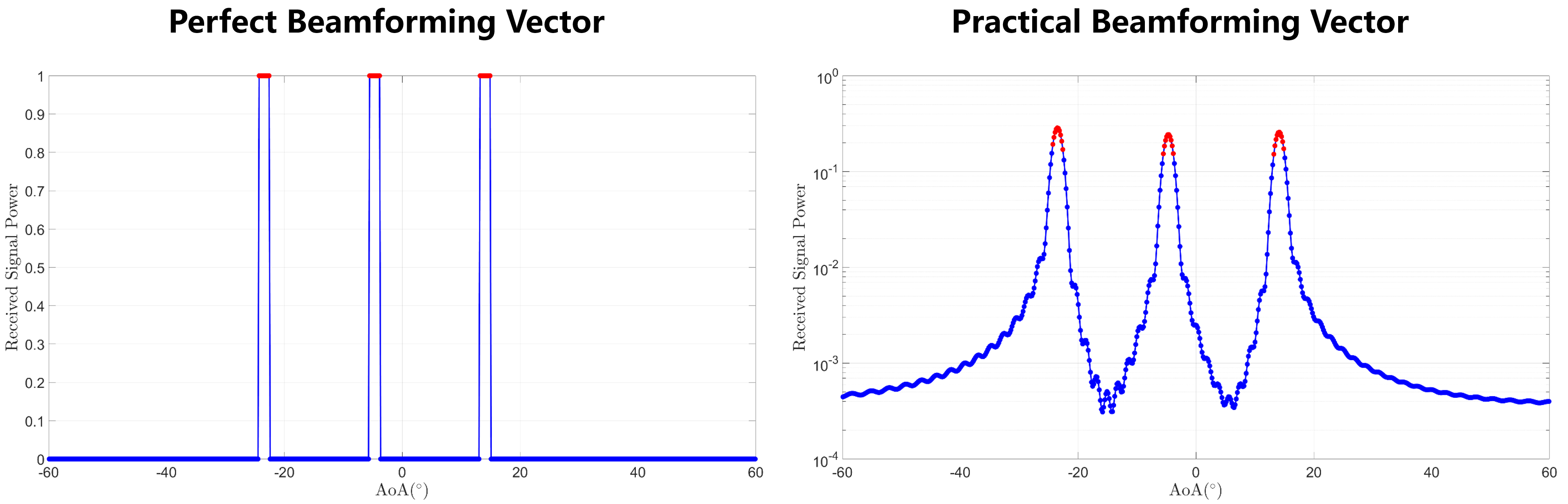}
\caption{The comparison of the power spectra generated by a perfect beamforming vector and a practical beamforming vector}
\label{perfect_practical}
\end{figure}

The above adaptive BA algorithm shares similarities with the one proposed in~\cite[Section II and III]{chiu2019active}, yet both remains a critical issue that may hinder its application in practical communication systems. Specifically, in above adaptive BA algorithm, it is unclear how the mapping function $\Gamma_\rmb^{\prime}(\cdot)$ should be constructed. In \cite{chiu2019active}, the authors assumed an ideal mapping function that generates a perfect beamforming vector, as shown in Fig \ref{perfect_practical} . However, in practical, it is challenging to design a perfect mapping function and thus there exists mismatch between the ideal search regions and the actual search regions, which results in differences between desired responses and actual responses. Detailed discussions are provided in Section \ref{sec_mr}.

\section{Main Results: How to Obtain Feasible Beamforming Vector}
\label{sec_mr}

To address the infeasibility of perfect questioner in \cite{chiu2019active}, this section introduces two methods to obtain practical questioner that generates feasible beamforming vectors. Mathematically, training a skilled questioner is equivalent to constructing a suitable mapping function $\Gamma_\rmb(\cdot)$, which was not addressed in~\cite{chiu2019active}.

\subsection{Linear Weighted Sum}
\label{mr Linear Weighted Sum}

We first propose a method by using a linear weighted sum of the steering vectors within the search region to yield beamforming vectors. Fix the number of sub-intervals $M \in \bbN$ and the number of secondary sub-intervals $k \in \bbN$, as shown in Fig. \ref{subinterval}. Specifically, we partition the whole search region $[0,1]$ into $M$ sub-intervals and we further divide each sub-interval into $k$ secondary sub-intervals, select the midpoint of each secondary sub-interval, and generate the steering vector using its corresponding angle as in Eq. \eqref{steering_vector}. Finally, we take the normalized linear weighted sum of all steering vectors as the beamforming vector. For simplicity, the weight of each steering vector is assumed to be 1.\footnote{It is possible that one can optimize the weight to achieve improved performance. However, this is beyond the scope of this paper since our goal is to propose a feasible BA algorithm using LWS and our simulation results show that this simple choice helps us achieve the goal.} The mapping function $\Gamma_\rmb^{\prime}(\cdot)$ corresponding to this method is defined as follows:

\begin{figure}[tb]
\centering
\includegraphics[width=0.6\columnwidth]{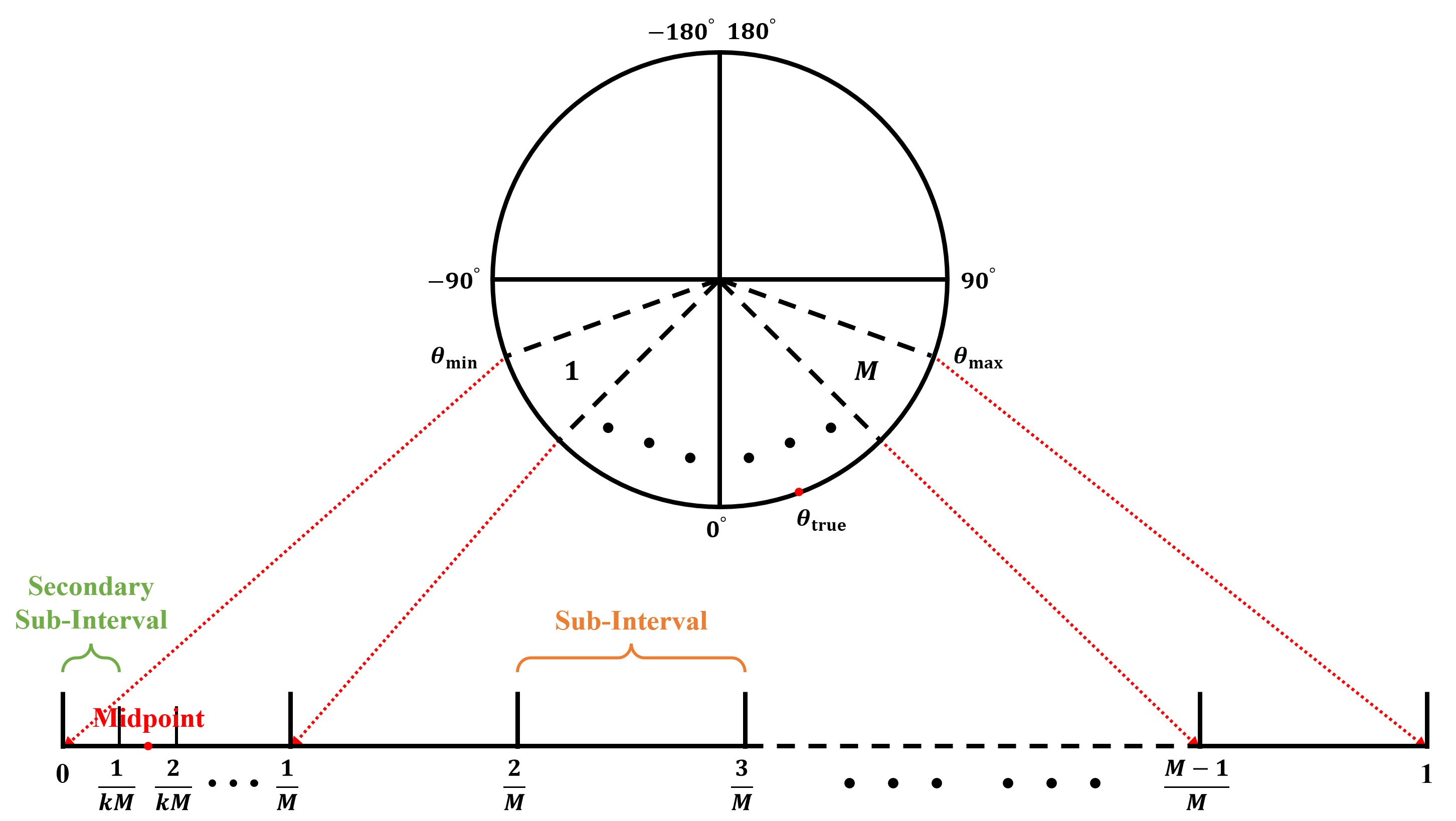}
\caption{Illustration of sub-intervals, secondary sub-intervals and midpoints of each sub-interval.}
\label{subinterval}
\end{figure}

\begin{definition} \label{def_LWS}
For any search region $\calA \subseteq [0,1]$, the mapping function $\Gamma_\rmb^{\prime}(\cdot)$ using linear weighted sum is defined as follows:
\begin{align}
\Gamma_\rmb^{\prime}(\calA) := \frac{\sum_{\calC_i \subseteq \calA} \sum_{k^{\prime} \in [k]} \bA(\theta_{\calC_i,k^{\prime}})}{\|\sum_{\calC_i \subseteq \calA} \sum_{k^{\prime} \in [k]} \bA(\theta_{\calC_i,k^{\prime}})\|_2}, \label{Gamma_b}
\end{align}
where $\theta_{\calC_i,k^{\prime}}$ denotes the angle corresponding to the midpoint of the $k^{\prime}$-th equidistant secondary sub-interval of the sub-interval $\calC_i$.
\end{definition}

With the available mapping function from search regions to beamforming vectors proposed in Definition \ref{def_LWS}, now we can provide a feasible adaptive BA algorithm, summarized in Algorithm \ref{alg_lws_1bit}. To evaluate the performance of the methods of LWS, we design numerical simulations with 1-bit measurement rule. At each time point $t$, we calculate the power of each received signal coming from the angle corresponding to the midpoint of the secondary sub-interval within the search region $\calA_t$. Specifically, we define the derived variable $K:=|\calA_t|M$, which denotes the number of sub-intervals within the search region $\calA_t$. For any search region $\calA_t \subseteq [0,1]$ and $k^{\prime} \in [K \times k]$, let $\theta_{k^{\prime}}$ denote the angle corresponding to the midpoint of the $k^{\prime}$-th secondary sub-interval within the search region, and let $\by_{\theta_{k^{\prime}}}^{\mathrm{q}}$ denote the received signal from the angle $\theta_{k^{\prime}}$. The definition of $|\by_{\mathrm{min}}^{\mathrm{q}}|^2$ is provided as follows:
\begin{align}
|\by_{\mathrm{min}}^{\mathrm{q}}|^2 := \min_{k^{\prime} \in [K \times k]} \big|\by_{\theta_{k^{\prime}}}^{\mathrm{q}}\big|^2. \label{y_min}
\end{align}

Unless otherwise specified, in all subsequent numerical simulations, we select the minimum $|\by_{\mathrm{min}}^{\mathrm{q}}|^2$ as the power threshold $\iota_t$. We calculate the power of each received signal coming from the angle corresponding to the midpoint of the secondary sub-interval within the non-search region and record the maximum $|\by_{\mathrm{max}}^{\mathrm{nonq}}|^2$, whose definition is similar to $|\by_{\mathrm{min}}^{\mathrm{q}}|^2$ in \eqref{y_min} as follows:
\begin{align}
|\by_{\mathrm{max}}^{\mathrm{nonq}}|^2 := \min_{k^{\prime} \in [(M-K) \times k]} \big|\by_{\theta_{k^{\prime}}}^{\mathrm{nonq}}\big|^2, \label{y_max}
\end{align}
where $\by_{\theta_{k^{\prime}}}^{\mathrm{nonq}}$ denote the received signal from the angle $\theta_{k^{\prime}}$ corresponding to the midpoint of the $k^{\prime}$-th secondary sub-interval within the non-search region $\bar{\calA}_t$. We illustrate $|\by_{\mathrm{min}}^{\mathrm{q}}|^2$ and $|\by_{\mathrm{max}}^{\mathrm{nonq}}|^2$ in Fig. \ref{power_spec_lws_overall}.

\begin{algorithm}[tb]
\caption{Adaptive Beam Alignment Algorithm Based on LWS with 1-bit Measurement Rule}
\label{alg_lws_1bit}
\begin{algorithmic}[0]
\REQUIRE
The number of sub-intervals $M \in \bbN$, the maximum number of time points $n \in \bbN$ and the acceptable error probability $\varepsilon \in (0,1)$.
\ENSURE
A estimate of AoA $\hat{\theta} \in [\theta_{\mathrm{min}},\theta_{\mathrm{max}}]$ and the number of time points $\tau \in [n]$.
~\\
\hrule 
~\\
\STATE Divide the angle range $[\theta_{\mathrm{min}},\theta_{\mathrm{max}}]$ into $M$ equally-sized non-overlapping sub-intervals $(\calC_1,\ldots,\calC_{M})$.
\STATE $\bm{\rho}_1 \leftarrow \{\frac{1}{M}\}_{1 \times M}, t \leftarrow 1$.
\WHILE{$t \leq n$}
\STATE \textbf{\emph{$\#$ Beamforming vector generation:}}
\STATE Sort the posterior probability vector $\bm{\rho}_t$ in descending order and get the sorted posterior probability vector $\bm{\rho}_t^{\downarrow}$.
\STATE Form a query $\calA_t$ based on $\bm{\rho}_t^{\downarrow}$ and convert it into a beamforming vector $\bw_t$ with a mapping function $\Gamma_\rmb^{\prime}(\cdot)$ in \eqref{Gamma_b}.
\STATE \textbf{\emph{$\#$ Signal sampling:}}
\STATE Receive a noisy signal $\by_t$ with $\bw_t$.
\STATE Convert $\by_t$ into a noisy response with a power threshold $\iota_t$.
\STATE \textbf{\emph{$\#$ Posterior update:}}
\STATE Update the posterior probability vector $\bm{\rho}_{t+1}$ by combining \eqref{update}, \eqref{Delta} and \eqref{z_t}.
\IF{$\max_{i \in [M]} \rho_{t+1,i} > 1-\varepsilon$}
\STATE Break.
\ELSE
\STATE $t \leftarrow t+1$
\ENDIF
\ENDWHILE
\STATE \textbf{\emph{$\#$ Decoding:}}
\STATE $\hat{i} \leftarrow \argmax_{i \in [M]} \rho_{t+1,i}$.
\STATE $\hat{\theta} \leftarrow \Gamma_\rmu^{-1}\big(\frac{2\hat{i}-1}{2M}\big),\tau \leftarrow t$.
\RETURN A estimate of AoA $\hat{\theta}$ and the number of time points $\tau$.
\end{algorithmic}
\end{algorithm}

Unless otherwise specified, in all subsequent numerical simulations, we fix the number of sub-intervals at $M=64$, the number of antennas at $N_R=64$, the number of secondary sub-intervals at $k=10$, and the maximum number of time points at $n=10$. We then plot the gap between $|\by_{\mathrm{max}}^{\mathrm{nonq}}|^2$ and $|\by_{\mathrm{min}}^{\mathrm{q}}|^2$ for various values of $K$ in Fig. \ref{1bit_power} (red curve with circles). Ideally, $|\by_{\mathrm{max}}^{\mathrm{nonq}}|^2$ should always be less than $|\by_{\mathrm{min}}^{\mathrm{q}}|^2$, which means that the accuracy of distinguishing whether there are signals from the search region based on this threshold should approach 1. However, it is worth noting that the curve of LWS in Fig. \ref{1bit_power} lies entirely above the baseline (black dashed line), which means even in the noise-free case, the measurement results of the received signals may be incorrect due to the deviation between the desired search regions and the actual search regions caused by the beamforming vectors generated by the imperfect questioner. Therefore, we would like to know if there is a better way to train questioners.

\subsection{Deep Neural Network} \label{mr Deep Neural Network}

In~\cite{sohrabi2021deep}, the authors proposed an end-to-end DNN that jointly performs beamforming vector design, posterior probability update, and AoA estimation (see Fig. 4 in~\cite{sohrabi2021deep}). As a result, the active-learning-based beam alignment algorithm is encapsulated within the ``black box'', making it challenging to analyze and improve its performance. To address this problem, instead of using DNN from end-to-end in all steps of BA, we only use DNN to design beamforming vectors. At each time point $t$, after choosing the current query $\calA_t$ based on the posterior probability vector $\bm{\rho}_t$, we generate the set of steering vectors in the same way as LWS in Section \ref{mr Linear Weighted Sum}. Subsequently, we use the set of generated steering vectors as the input to a pre-trained DNN, which serves as the mapping function $\Gamma_\rmb^{\prime}(\cdot)$, and obtain the corresponding beamforming vector as output. The mapping function $\Gamma_\rmb^{\prime}(\cdot)$ corresponding to the DNN method is defined as follows:

\begin{definition} \label{def_DL}
Given any number of secondary sub-intervals $k \in \bbN$ for each sub-interval, for any search region $\calA \subseteq [0,1]$, the mapping function $\Gamma_\rmb^{\prime}(\cdot)$ is designed by a multi-layer fully connected DNN that consists of:
\begin{itemize}
\item an input layer: Receiving the real part $\frakR$ and the imaginary part $\frakI$ of steering vectors of angles corresponding to the midpoint of $k$ secondary sub-intervals of each sub-interval included in the query $\calA$.
\item multiple hidden layers: Comprising trainable weights $\frakA$ and biases $\frakB$ of the DNN, as well as activation functions $\zeta(\cdot)$, for each layer.
\item an output layer: Generating the beamforming vector $\bw$ as follows:
\begin{align}
\bw = \zeta_L\bigg( \frakA_L \zeta_{L-1}\Big( \cdots \zeta_1\big(\frakA_1[\frakR,\frakI]+\frakB_1\big) \cdots \Big)+\frakB_L \bigg) ,
\end{align}
where $L \in \bbN$ denotes the number of layers in DNN.
\end{itemize}
\end{definition}

\begin{figure}[tb]
\centering
\includegraphics[width=0.6\columnwidth]{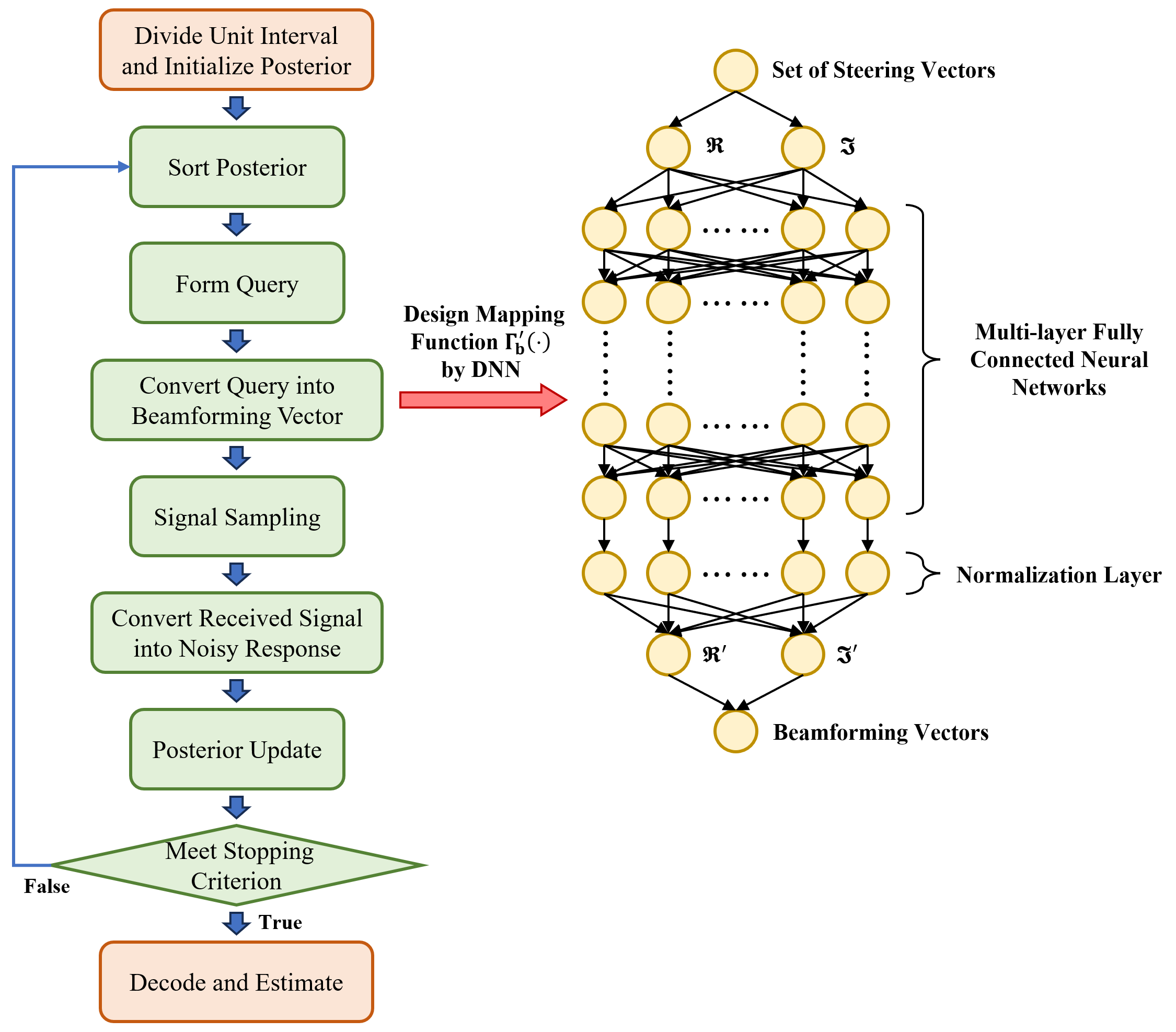}
\caption{The block diagram of the DNN-based adaptive BA algorithm.}
\label{DNN}
\end{figure}

Except for the mapping function $\Gamma_\rmb^{\prime}(\cdot)$, the algorithm of adaptive BA algorithm based on LWS and the one based on DNN are identical. in Fig. \ref{DNN}, we provide a block diagram of the DNN-based adaptive BA algorithm.

\begin{figure}[tb]
\centering
\subfloat[The gap between the maximum power of received signals from the non-search region and the minimum power of received signals from the search region.]{
\centering
\includegraphics[width=0.4\textwidth]{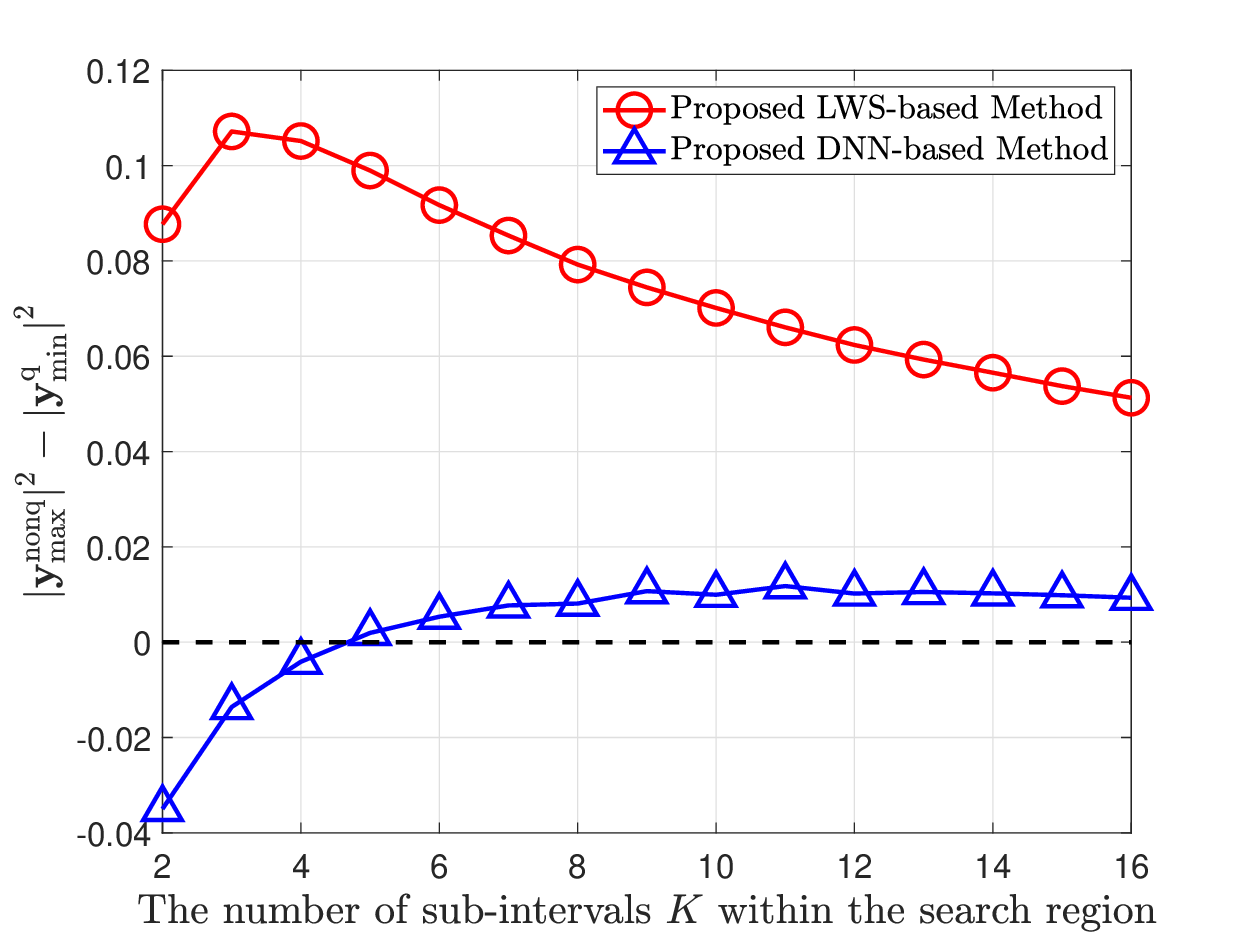}
\label{1bit_power}
}
\hfill
\subfloat[The accuracy that the actual response of each query during the adaptive BA algorithm matches the desired response.]{
\centering
\includegraphics[width=0.4\columnwidth]{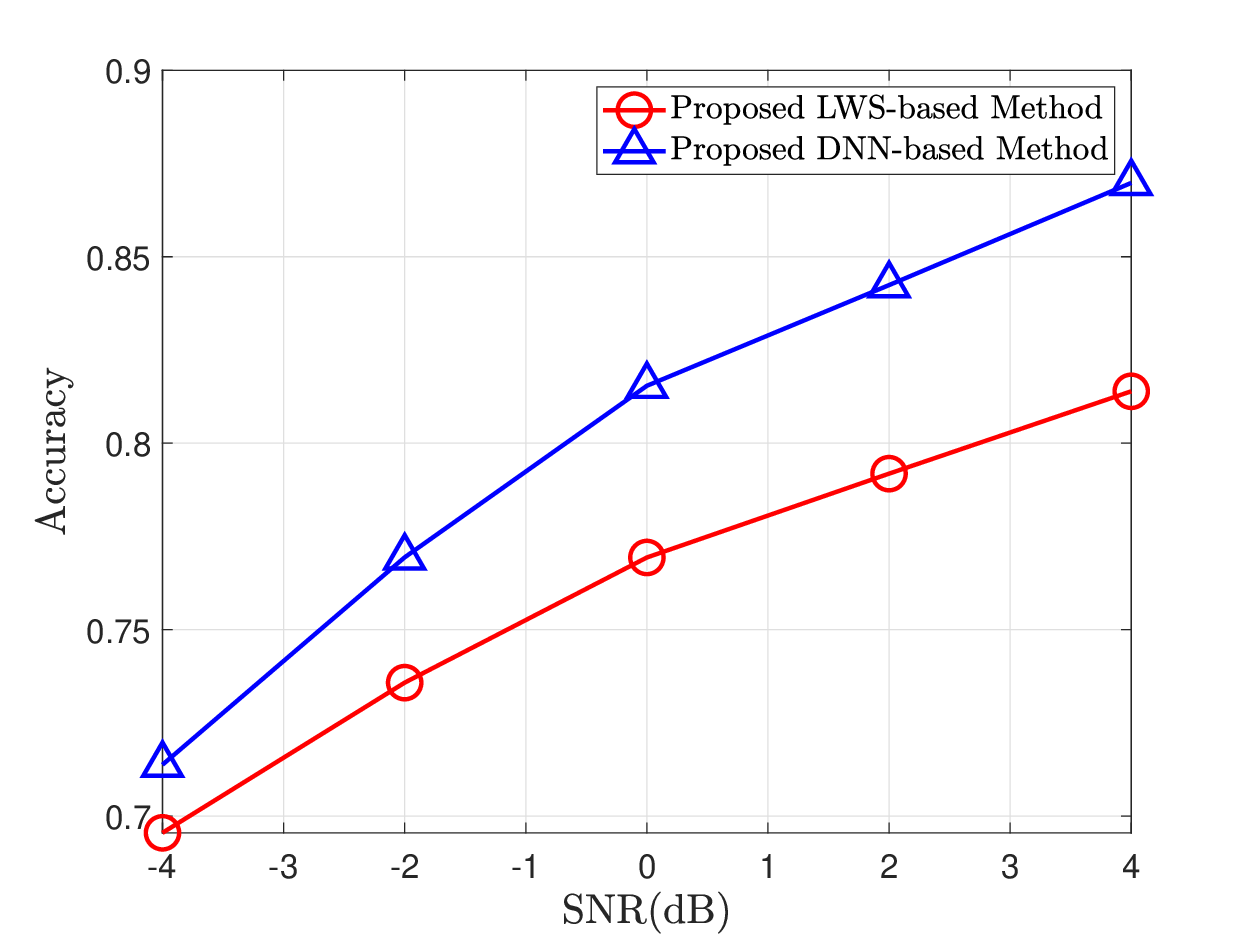}
\label{1bit_accuracy}
}
\caption{Performance comparison with 1-bit measurement rule between the LWS-based method and the DNN-based method. As observed, the DNN-based method achieves better performance than the LWS-based method under the 1-bit measurement rule.}
\label{lws_vs_dnn_1bit}
\end{figure}

Similarly to the LWS-based method in Section \ref{mr Linear Weighted Sum}, we plot the gap between $|\by_{\mathrm{max}}^{\mathrm{nonq}}|^2$ and $|\by_{\mathrm{min}}^{\mathrm{q}}|^2$ for various values of $K$ using the DNN-based method in Fig. \ref{1bit_power} (blue curve with triangles). Note that when $K$ is small ($K \leq 4$), the blue curve is below the black baseline, meaning that at this moment the questioner trained by DNN is perfect. Although the performance of the questioner degrades as $K$ increases, it is still significantly better than the questioner trained by LWS, as it is notably closer to the baseline.

Furthermore, to more intuitively demonstrate the performance comparison between the two methods, we plot the average accuracy under different signal-to-noise ratio (SNR) conditions, where the accuracy is defined as the average probability that the actual response of each query during the adaptive BA algorithm matches the desired response, as shown in Fig. \ref{1bit_accuracy}. After running $1 \times 10^3$ times under the query-independent AWGN channel, the simulation results show that the DNN-based method (blue curve with triangles) can achieve higher accuracy than the LWS-based method (red curve with circles) at the same SNR. Furthermore, the performance gap between the two methods becomes significantly larger as SNR increases. This is because there are two factors contributing to mismatches between the actual measurement results and the desired measurement results, i.e., channel noise and imperfectly trained questioner. When channel conditions improve, the impact of the latter becomes more pronounced.

\subsection{Performance Analysis under the 1-bit Measurement Rule} \label{mr Analysis of Two Methods with 1-bit Measurement Rule}

\begin{figure}[tb]
\centering
\subfloat[Overall diagram.]{
\centering
\includegraphics[width=0.6\textwidth]{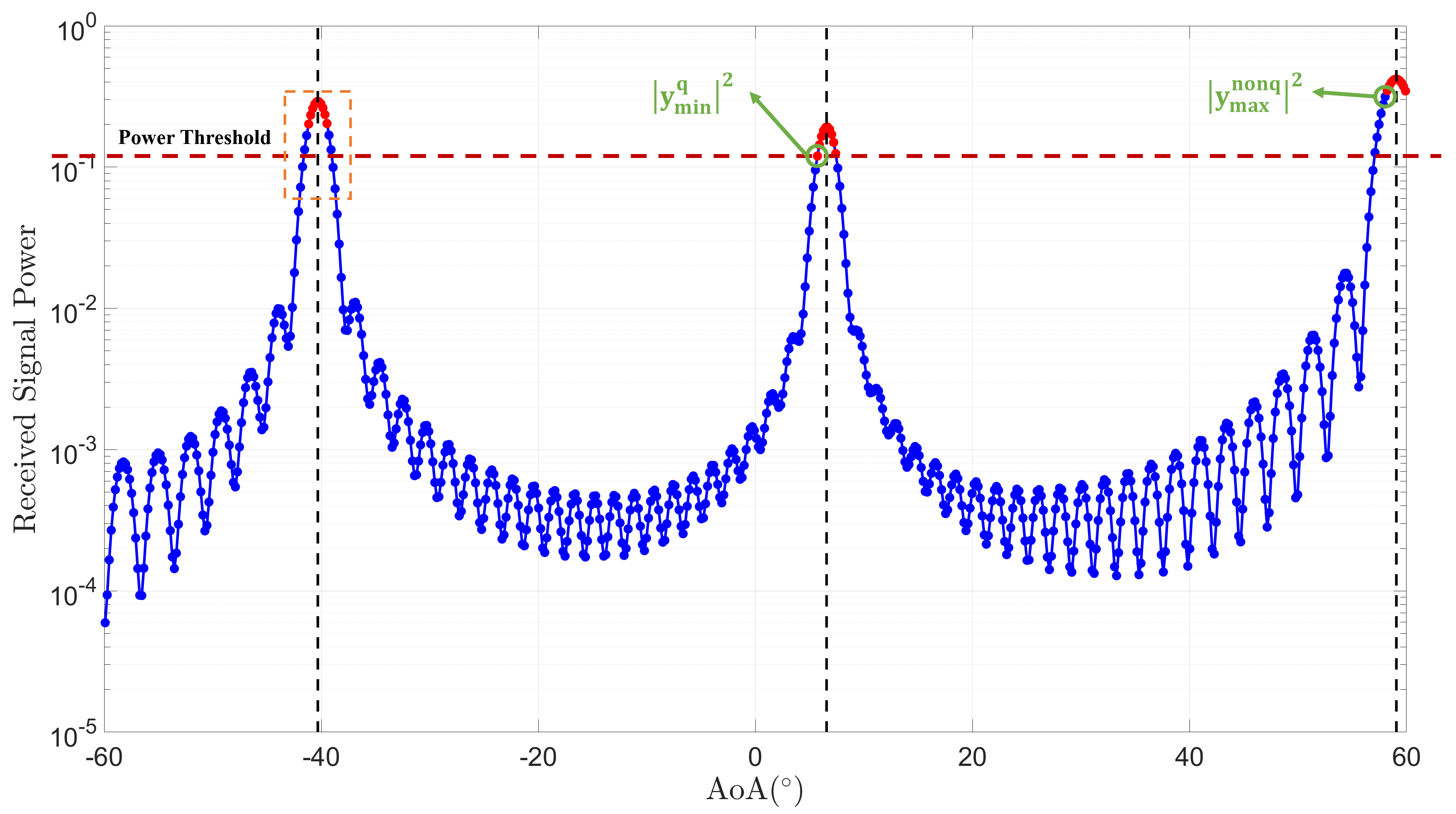}
\label{power_spec_lws_overall}
}
\\
\subfloat[Enlarged diagram of the orange dashed box.]{
\centering
\includegraphics[width=0.6\columnwidth]{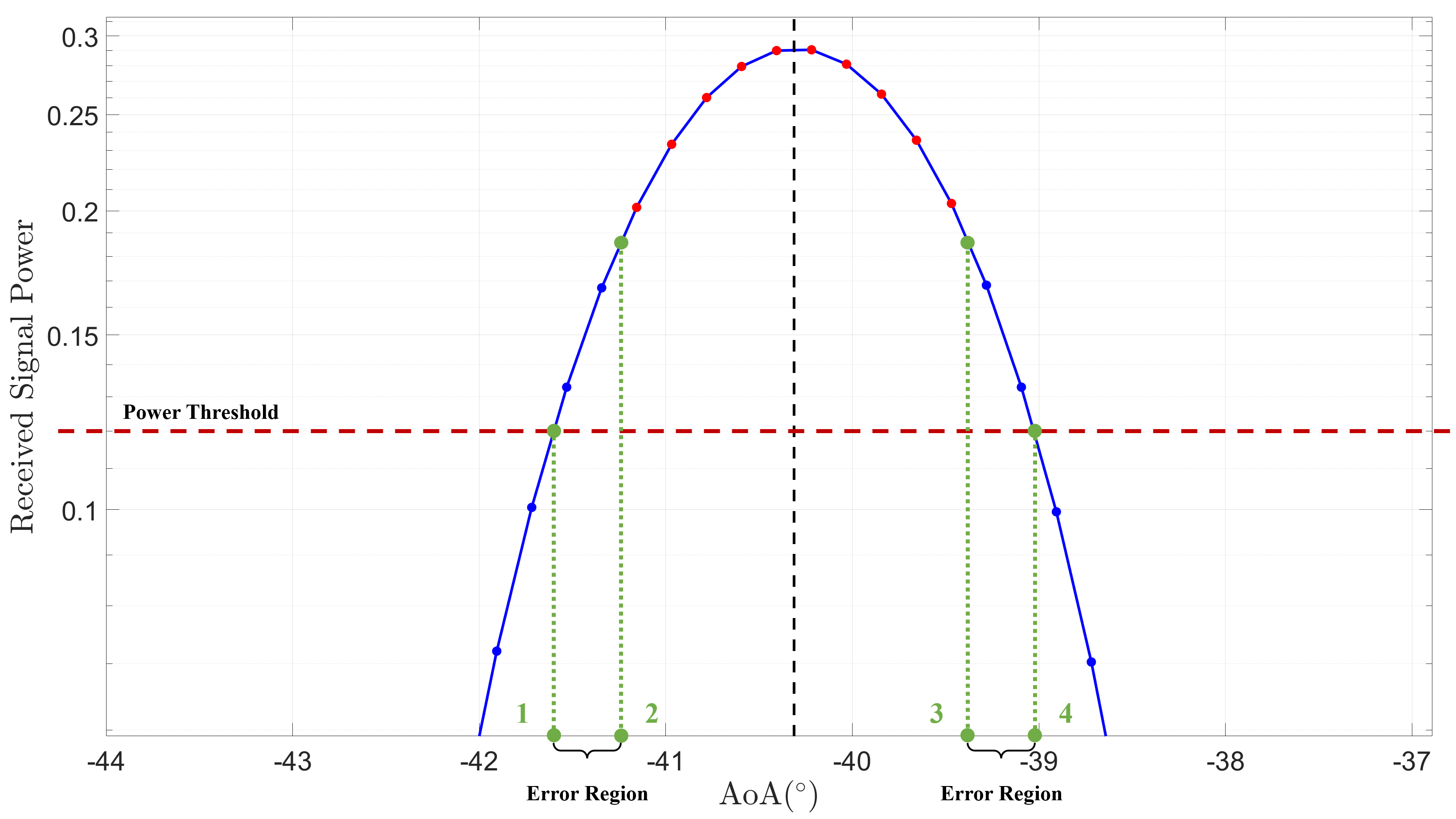}
\label{power_spec_lws_detail}
}
\caption{Power spectrum for a specific beamforming vector generated by a questioner trained using the LWS-based method with $K=3$ under the 1-bit measurement rule. As observed, for any beamforming vector generated by a practical mapping function under the 1-bit measurement rule, there will be an error region that may lead to the mismatches between actual measurement results of the received signals and the ideal result assumed in~\cite{chiu2019active}.}
\label{power_spec_lws}
\end{figure}

Recall the definition of 1-bit measurement rule in \eqref{1bit}, which serves as one of the main assumptions in the analysis and simulations of~\cite{chiu2019active}. In this subsection, we illustrate what an imperfectly trained questioner is and why it leads to mismatches between the actual measurement results and the desired measurement results. Set the number of sub-intervals within the search region as $K=3$. Then we use the LWS-based method to generate a beamforming vector, and plot the power spectrum of signals received from different directions using this beamforming vector, as shown in Fig. \ref{power_spec_lws_overall}. Specifically, three black vertical dashed lines represent the angles corresponding to the respective midpoints of the three sub-intervals that the beamforming vector intends to search. The red dots represent the power of received signals from the search region, while the blue dots represent the power of received signals from the non-search region. The red horizontal dashed line represents the power threshold selected for this moment. When the trained questioner is perfect, there are only red dots above the red horizontal dashed line and only blue dots below it. Fig. \ref{power_spec_lws_detail} is an enlargement of the orange dashed box in Fig. \ref{power_spec_lws_overall}, from which it can be seen that some blue dots are also above the red horizontal dashed line, except for the red dots. Currently, the directions between green dots 1 and 2, as well as between green dots $3$ and $4$, are deemed to be error regions, meaning that signals from these regions (non-search region) will be mistaken for signals from the search region, thereby leading to the mismatches of measurement results.

\begin{figure}[tb]
\centering
\includegraphics[width=0.6\columnwidth]{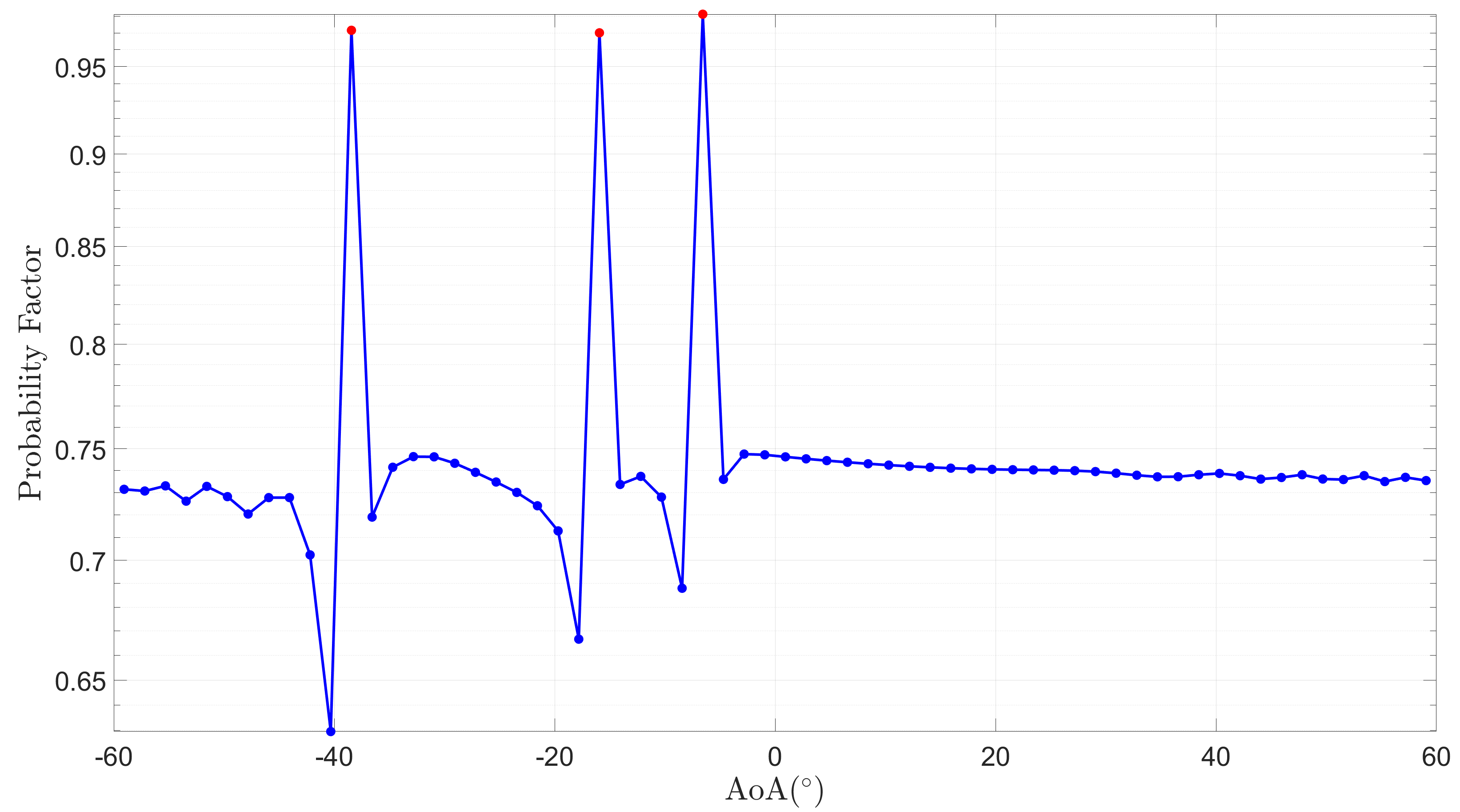}
\caption{The spectrum of the probability factor for a specific beamforming vector generated by a questioner trained using the LWS-based method with $K=3$ under the full measurement rule. As observed, the three peaks of the spectrum plotted for the beamforming vector generated by the DNN-based method are narrower in width than those generated by the LWS-based method in Fig \ref{power_spec_lws_overall}.}
\label{probFactor_spec}
\end{figure}

\begin{figure}[tb]
\centering
\subfloat[Power spectrum of the LWS-based method.]{
\centering
\includegraphics[width=0.6\textwidth]{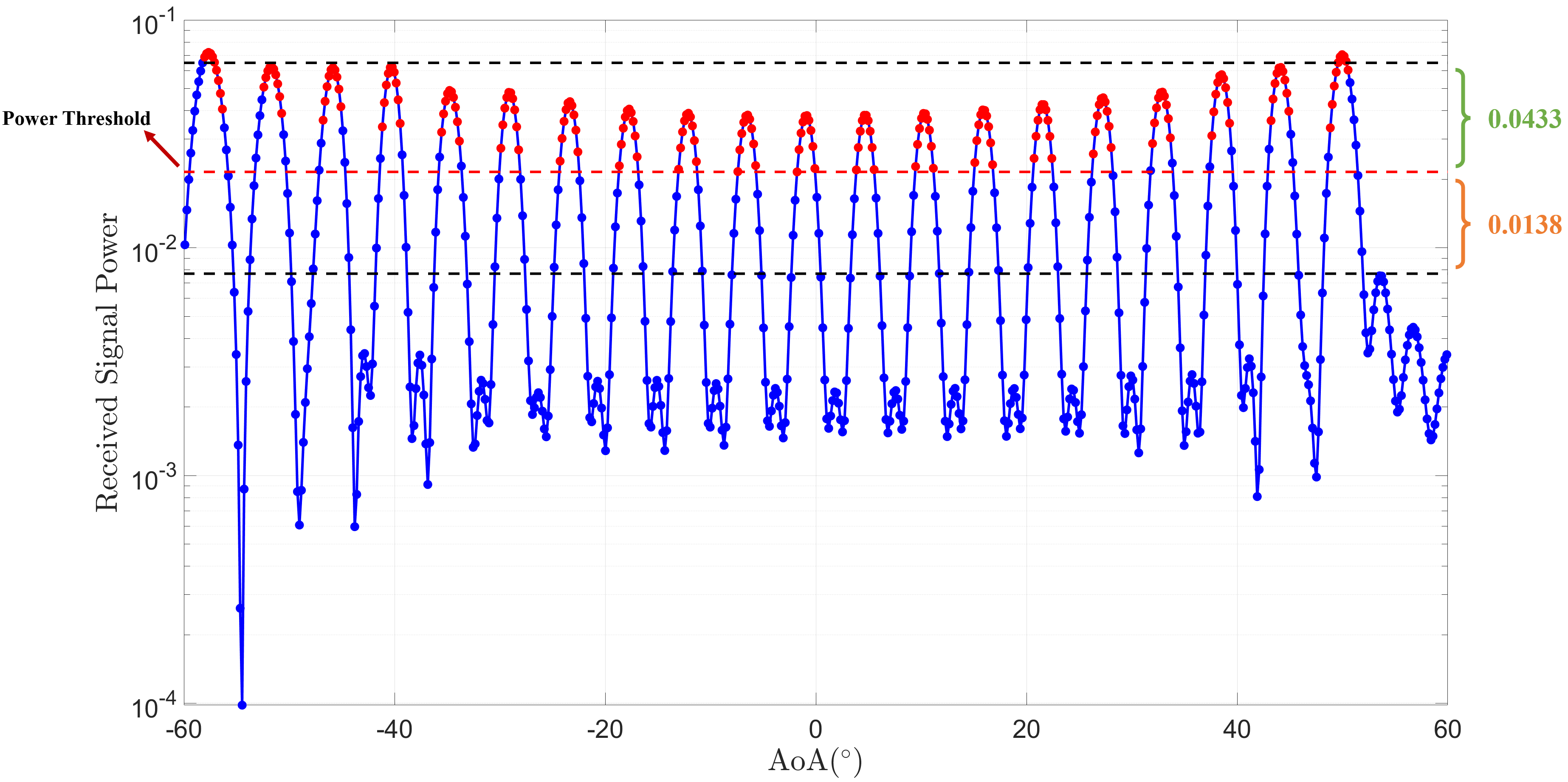}
\label{power_spec_k20_lws}
}
\\
\subfloat[Power spectrum of the DNN-based method.]{
\centering
\includegraphics[width=0.6\columnwidth]{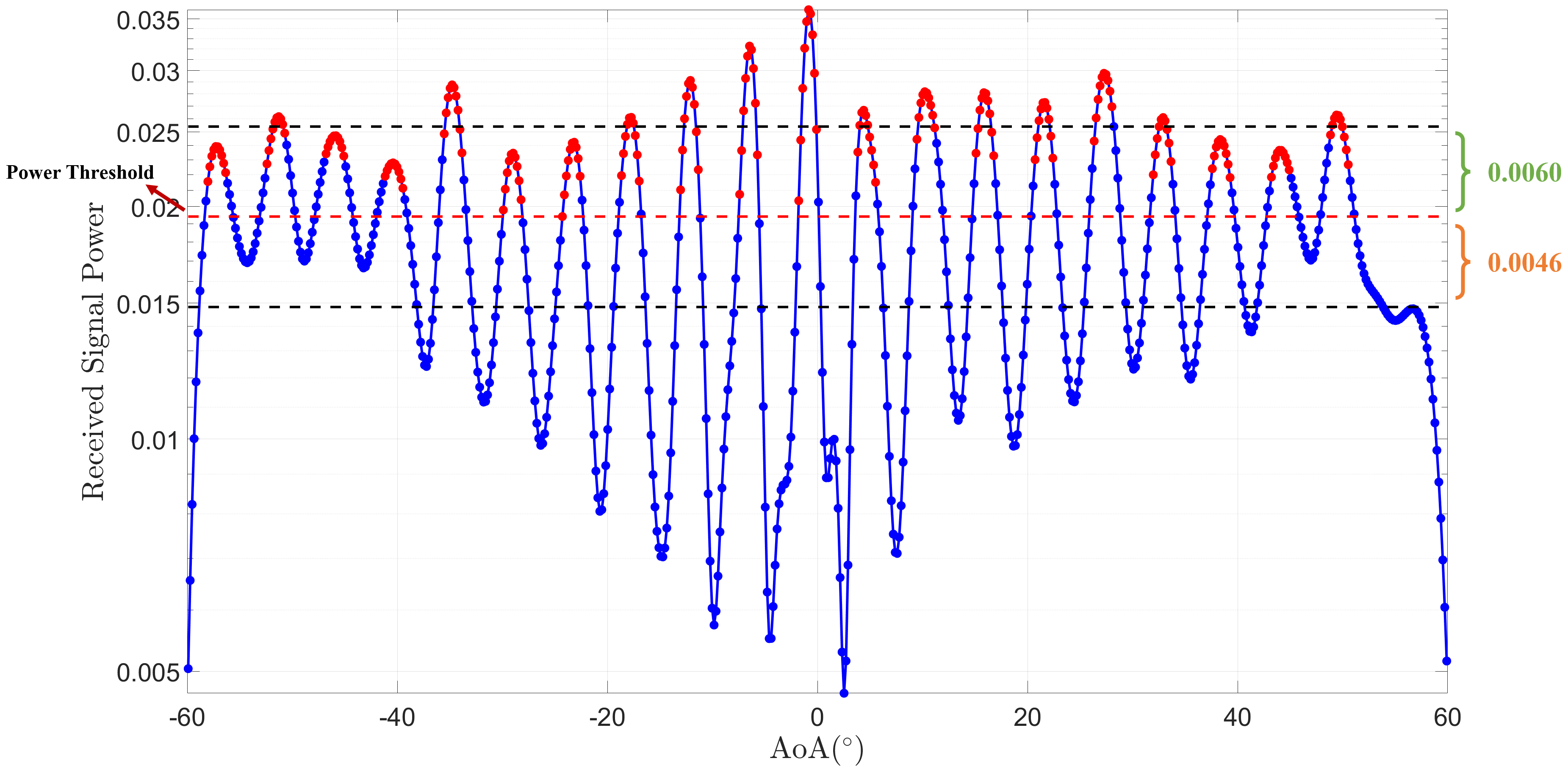}
\label{power_spec_k20_dnn}
}
\caption{Power spectra for two beamforming vector generated respectively by questioners trained using the LWS-based method and the DNN-based method for a same query with $K=20$ under the 1-bit measurement rule. As observed, when $K$ is relatively large, the power spectrum for the beamforming vector generated by the DNN-based method is flatter than those generated by the LWS-based method, implying that DNN-based method is inferior to the LWS-based method in terms of noise resistance.}
\label{power_spec_k20}
\end{figure}

\begin{figure}[tb]
\centering
\includegraphics[width=0.6\columnwidth]{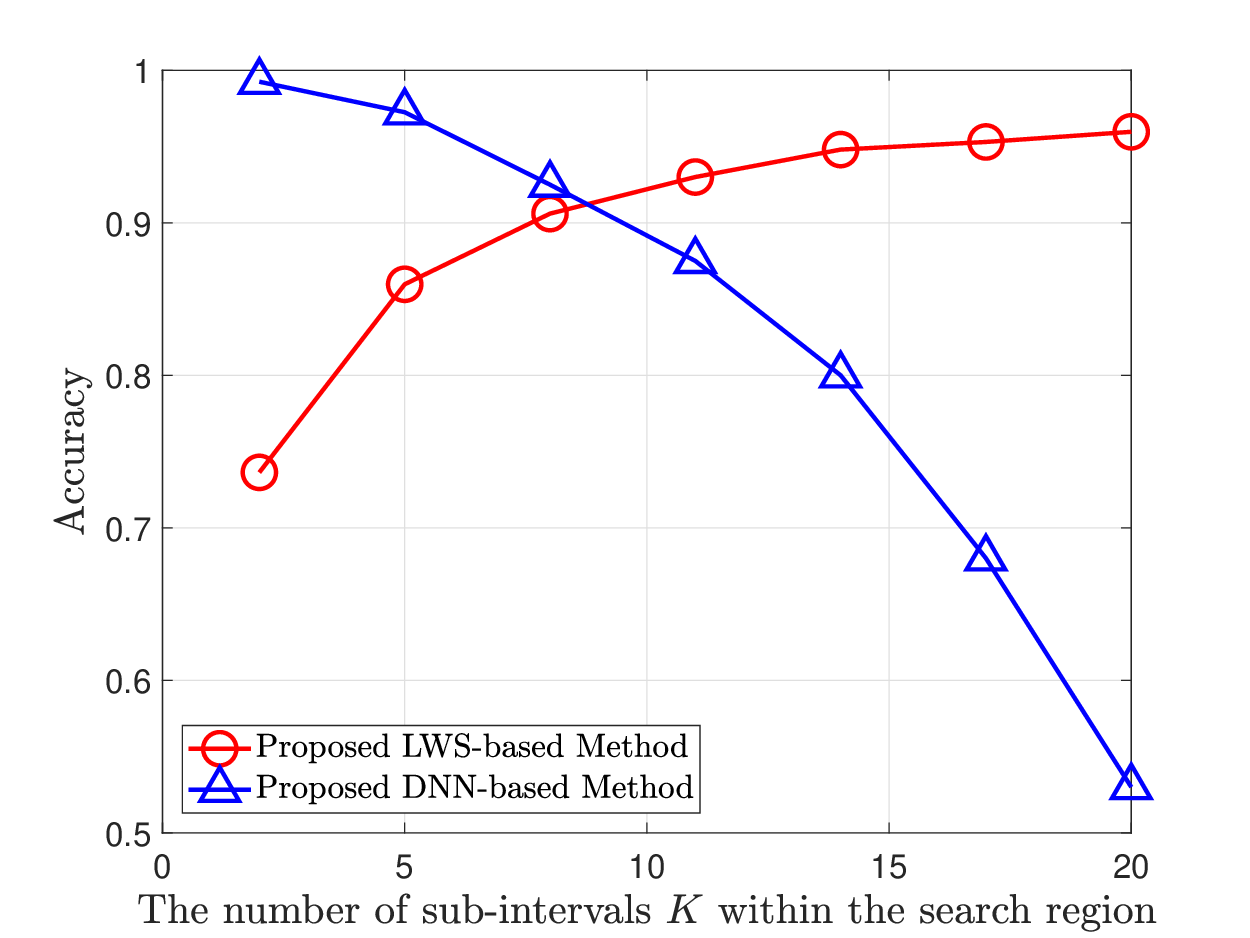}
\caption{The accuracy of the selected power threshold with 1-bit measurement rule when $\mathrm{SNR}=0~\mathrm{dB}$. As observed, when $K$ increases, the accuracy of the selected power threshold with the beamforming vector generated by the LWS-based method also increases, while the DNN-based method shows the opposite trend.}

\label{1bit_accuracy_large}
\end{figure}

However, the DNN-based method also has a drawback. When the number of sub-intervals within the search region is large, compared with the LWS-based method, the DNN-based method is prone to flattening the entire power spectrum, as shown in Fig. \ref{power_spec_k20}. The search region is set as $\calA = \cup_{j \in [2:3:59]} \calC_j$, so that the number of sub-intervals within the search region is $K = 20$. For the LWS-based method, the gap between $|\by_{\mathrm{max}}^{\mathrm{nonq}}|^2$ (upper black horizontal dashed line) and $|\by_{\mathrm{min}}^{\mathrm{q}}|^2$ (middle red horizontal dashed line) is $0.0433$ (green number in Fig. \ref{power_spec_k20_lws}). While for the DNN-based method, the gap is $0.0060$ (green number in Fig. \ref{power_spec_k20_dnn}). Thus the gap for the DNN-based method is less than that for the LWS-based method, where the magnitude relationship is consistent with that in Fig. \ref{1bit_power}. However, we note that for the DNN-based method, the gap between $|\by_{\mathrm{min}}^{\mathrm{q}}|^2$ and the average power of each received signal from the non-search region (lower black horizontal dashed line) is $0.0046$ (orange number in Fig. \ref{power_spec_k20_lws}), which is also less than the gap of $0.0138$ (orange number in Fig. \ref{power_spec_k20_dnn}) for the LWS-based method. Due to this characteristic, the ability of the beamforming vector generated by the DNN-based method to resist noise interference is weaker than that generated by the LWS-based method, which is intuitively reflected in their power spectra: the power spectrum of the former is overall flatter in Fig. \ref{power_spec_k20}. To further demonstrate the impact of the flattened power spectrum, we plot the accurary for various values of $K$ in Fig. \ref{1bit_accuracy_large}. Note that when the number of sub-intervals $K$ within the search region is relatively large, compared with the LWS-based method (red curve with circles), the accuracy of the selected power threshold in the DNN-based method (blue curve with triangles) decreases rapidly with the increase of $K$ with $\mathrm{SNR}=0~\mathrm{dB}$\footnote{Note that there is an intersection point between the accuracy of the two methods, which means that the entire adaptive BA algorithm can achieve the best performance when we use the DNN-based method on the left side of the intersection point and the LWS-based method on the right side. The location of the intersection point may be influenced by many factors, such as the number of sub-intervals and SNR, which is worth analyzing in subsequent research.}.

\subsection{Performance Analysis under the Full Measurement Rule}
\label{mr Full Measurement Rule}

To investigate the performance limits of the two methods, we further study the adaptive BA algorithm with full measurement rule, where $\by_t$ is directly used as the response to the query $\calA_t$ at each time point $t$.

For the DNN-based method, we first would like to know whether the questioners trained by NNs with 1-bit measurement rule can also achieve good performance with full measurement rule, just like the LWS-based method. Recall the definitions of the posteriors update in \eqref{update} and the function $\Delta(\cdot)$ in \eqref{Delta}, we can compute the posterior probability vector $\bm{\rho}_{t+1}$ as follows:
\begin{align}
\rho_{t+1,i}:=\frac{\rho_{t,i}\nu(\theta_i,\by_t,\bw_t)}{\sum_{i^{\prime} \in [M]}\rho_{t,i^{\prime}}\nu(\theta_{i^{\prime}},\by_t,\bw_t)}, \forall~i \in [M], \label{update_full}
\end{align}
where
\begin{align}
\nu(\theta_i,\by_t,\bw_t) := e^{-|\by_t-\bw_t\bA(\theta_i)|^2}. \label{nu}
\end{align}

\begin{figure}[tb]
\centering
\includegraphics[width=0.6\columnwidth]{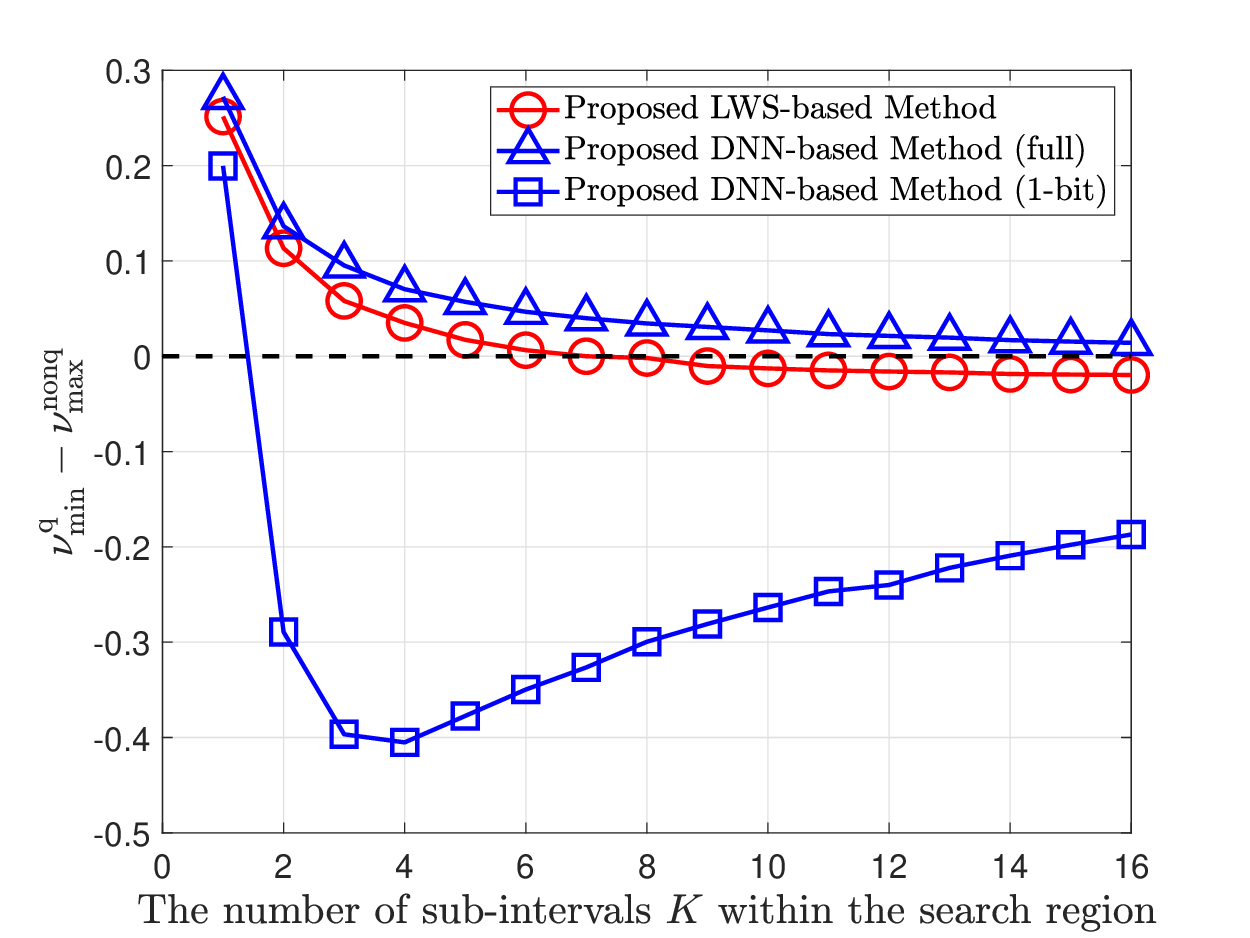}
\caption{The gap between $\nu_{\mathrm{max}}^{\mathrm{nonq}}$ and $\nu_{\mathrm{min}}^{\mathrm{q}}$ with three methods for training the questioners. As observed, the DNN-based method achieves better performance than the LWS-based method under the full measurement rule. Furthermore, for DNN-based methods, the trained mapping function is different under different measurement rules.}
\label{full_exp}
\end{figure}

To evaluate the performance with full measurement rule, at each time point $t$, we calculate the probability factor $\nu(\theta^{\mathrm{q}},\by_t,\bw_t)$ for each angle $\theta^{\mathrm{q}}$ corresponding to the midpoint of the sub-interval within the search region $\calA_t$ and select the minimum $\nu_{\mathrm{min}}^{\mathrm{q}}$. Furthermore, we calculate $\nu(\theta^{\mathrm{nonq}},\by_t,\bw_t)$ for each angle $\theta^{\mathrm{nonq}}$ corresponding to the midpoint of the sub-interval within the non-search region $\bar{\calA}_t$ and select the maximum $\nu_{\mathrm{max}}^{\mathrm{nonq}}$. Similarly, we plot the gap between $\nu_{\mathrm{max}}^{\mathrm{nonq}}$ and $\nu_{\mathrm{min}}^{\mathrm{q}}$ for various values of $K$ in Fig. \ref{full_exp}. Ideally, $\nu_{\mathrm{max}}^{\mathrm{nonq}}$ should always be less than $\nu_{\mathrm{min}}^{\mathrm{q}}$, and the larger the margin by which $\nu_{\mathrm{max}}^{\mathrm{nonq}}$ is less than $\nu_{\mathrm{min}}^{\mathrm{q}}$, the more rapidly and accurately the adaptive BA algorithm can be completed. However, it is worth noting that the curve of the DNN-based method with 1-bit measurement rule (blue curve with squares) in Fig. \ref{full_exp} lies almost below the baseline (black dashed line), and its gap with the baseline is significantly larger than that of the LWS-based method (red curve with circles).

Therefore, we retrained the NNs and plot the gap between $\nu_{\mathrm{max}}^{\mathrm{nonq}}$ and $\nu_{\mathrm{min}}^{\mathrm{q}}$ in Fig. \ref{full_exp}. Note that the blue curve with triangles is always above the black baseline, meaning that the questioners trained by the DNN-based method is perfect. Compared with the NNs trained with 1-bit measurement rule, the change in the optimization problem leads to a significant improvement in the performance of the questioners.

In summary, with the above methods for training questioner, now we can address the critical issue of the lack of the feasible and interpretable methods to convert the queries into beamforming vectors. In the next section, we will demonstrate the performance of the adaptive BA algorithms through system simulations.

\section{Numerical Illustration of Superior Performance of Our BA Algorithms} 
\label{sec_nr}

In the above sections, we propose two methods for training skilled questioners to generate the beamforming vectors and analyze the performance gaps with both 1-bit measurement rule and full measurement rule. However, all analysis and simulations are conducted from a local perspective. In this section, we illustrate the performance of the proposed adaptive BA algorithms in practical communication systems, compared with several existing schemes in the literature. Before presenting the numerical results, we briefly explain why the query-dependent channel model is more suitable. This aspect also lacks detailed discussions and analyses in~\cite{chiu2019active}.

\subsection{Validity of Query-Dependent Channel Models} \label{nr Validity of Query-Dependent Channel Models}

The query-dependent channel model was proposed by Chiu and Javidi~\cite{chiu2016sequential}, inspired by a phenomenon that the noise can gradually accumulate as the query region expands in many practical applications such as target localization using sensor networks~\cite{tsiligkaridis2014collaborative}. Thus compared with the query-independent channel model~\cite{jedynak2012twenty,chung2017unequal}, the query-dependent~\cite{kaspi2017searching,lalitha2018improved,zhou2021resolution} channel model is a more practical channel model, which can be further confirmed by the proposed adaptive BA algorithm with both 1-bit measurement rule and full measurement rule.

Recall the definitions of query-dependent channels in \eqref{bsc} and \eqref{awgn}, where the BSC corresponds to the 1-bit measurement rule and the AWGN channel corresponds to full measurement rule. As mentioned in Section \ref{mr Deep Neural Network}, mismatches between the actual measurement results and the desired measurement results are caused by channel noise and imperfectly trained questioner. The former is generally regarded as AWGN in communication systems, while the latter is closely related to queries. As shown in Fig. \ref{1bit_power}, in the noise-free case, the gap between $|\by_{\mathrm{max}}^{\mathrm{nonq}}|^2$ and $|\by_{\mathrm{min}}^{\mathrm{q}}|^2$ varies with the number of sub-intervals $K$ within the search region, which means that the impact of imperfectly trained questioner varies with the size of queries, with 1-bit measurement rule. We can also draw the same conclusion with full measurement rule from Fig. \ref{full_exp}. From the above two figures, we have verified the rationality of the query-dependent channel model in the adaptive BA algorithm.

\subsection{Benchmark Algorithms} \label{nr Details of Benchmark Algorithms}

\subsubsection{Naive beam sweeping algorithm} In this simplest approach, we partition the unit interval $[0,1]$ into $n$ equal-sized non-overlapping sub-intervals based on the maximum number of samples $n$. Then we sequentially select the corresponding angle of the midpoint of each sub-interval to generate the steering vector, and take it as the beamforming vector. After $n$ samples, we find the sub-interval with the highest power of the received signal, and determine the angle corresponding to its midpoint as the estimate of AoA.

\subsubsection{Hierarchical beam sweeping algorithm proposed in~\cite{ZTE2017OnCSI}} 5G NR system incorporates the beam management procedure to acquire and maintain beams, one component of which is beam sweeping~\cite{li2020beam}. The beam sweeping algorithm uses a two-stage exhaustive search algorithm for both transmitting and receiving ends, which can reduce the number of queries under the same resolution limit. Specifically, we first partition the unit interval with a low resolution and locate the sub-interval where the AoA lies. Then we partition this sub-interval with a higher resolution and finally find the secondary sub-interval where AoA is located.

\subsubsection{Ideal hiePM-based strategy proposed in~\cite{chiu2019active}}

In fact, the authors provided a beam design method based on ideal assumptions in~\cite[Eq. (41)]{chiu2019active}. However, this approximate solution to~\cite[Eq. (38)]{chiu2019active} is flawed when some probing directions in a given query are too close to each other. Therefore, we slightly modify it to enable its use in simulations, as follows:
\begin{align}
\bw(\calD_l^k)=C_s(\bA_{\text{BS}}\bA_{\text{BS}}^H+\sigma_0^2\bI)^{-1}\bA_{\text{BS}}\bG_{\calD_l^k},
\end{align}
where $\calD_l^k$ denotes the probing region, $C_s$ denotes a normalization constant, $\bA_{\text{BS}}$ denotes a matrix of array manifolds, and $\bG_{\calD_l^k}$ denotes a vector indicating the probing region. The term $\sigma_0^2\bI$ we added here is for diagonal loading, where $\sigma_0^2$ denotes a normalization factor and $\bI$ denotes the identity matrix. Furthermore, the threshold selection of this strategy follows the approach described in~\cite[Appendix A]{chiu2019active}.

\subsubsection{DNN-based adaptive sensing strategy proposed in~\cite{sohrabi2021deep}}

This design adopts the framework of adaptive BA with full measurement rule and uses a DNN architecture to estimate the AoA. Specifically, the strategy takes the current posterior distribution together with the other available system parameters (e.g., power constraint $P$ and time point $t$) as inputs, and takes the estimate of AoA as output. It should be mentioned that most components of this strategy are covered in the DNN architecture, which is essentially different from that of this paper. Furthermore, this strategy discusses two scenarios separately: when the fading coefficient $\alpha$ is known and when it is unknown. Here, we only compare the first scenario.

\subsection{Simulation Results} \label{nr Simulation Results}

In this subsection, we can mainly draw two conclusions: i) Our proposed methods outperform all benchmark algorithms, and they outperform the best-performing benchmark algorithm by $8$ dB with the same average quadratic loss order under the same simulation settings. ii) Compared with the methods with 1-bit measurement rule, the methods with full measurement rule show better performance due to the fact that entire information of received signals can be available and utilized, which is consistent with the analysis and simulations in Section \ref{sec_mr}.

We first examine the performance of the proposed methods and the above benchmark algorithms with the range of angles $[\theta_{\mathrm{min}},\theta_{\mathrm{max}}] = [-60^{\circ},60^{\circ}]$. For the settings of training parameters of NNs, we implement the proposed NN on TensorFlow~\cite{abadi2016tensorflow} by employing Adam optimizer~\cite{kingma2014adam} with a learning rate $0.0001$. And for each size of queries, we consider $4$-layer neural networks with dense layers of widths $[2kKM, 1024, 1024, 2M]$. We also employ an appropriate normalization layer at the last layer. We consider $10$ batches per epoch, and terminate the training procedure when the performance for the validation set has not improved\footnote{The source code for this paper is available at: https://github.com/Chunsong-Sun/Adaptive-Beam-Alignment-using-Noisy-Twenty-Questions-Estimation-with-Trained-Questioner}.

\begin{figure}[tb]
\centering
\subfloat[Average quadratic loss versus SNR for the two proposed methods with 1-bit measurement rule, the ideal hiePM-based strategy, and the two beam sweeping algorithms.]{
\centering
\includegraphics[width=0.4\textwidth]{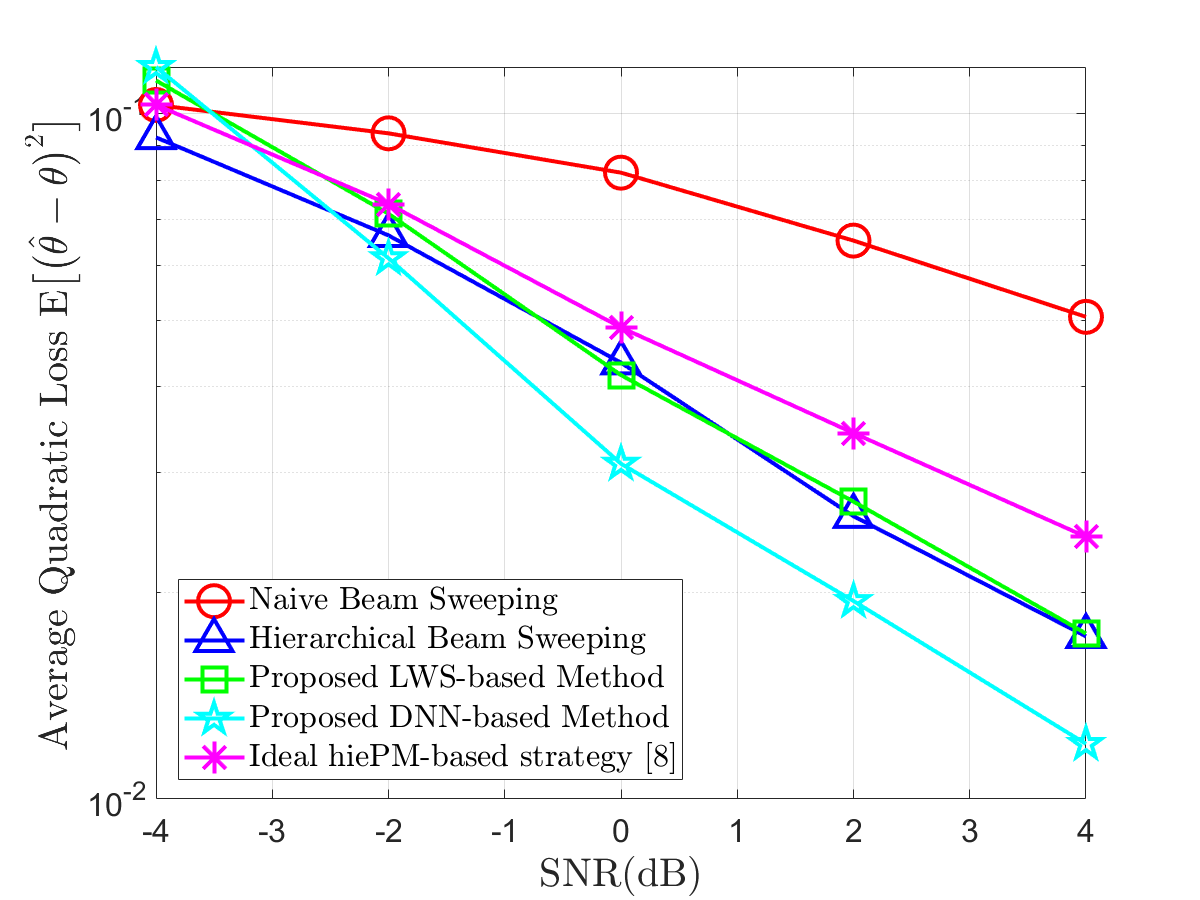}
\label{quadratic_loss_1bit}
}
\hfill
\subfloat[Average quadratic loss versus SNR for the two proposed methods with full measurement rule and the DNN-based adaptive sensing strategy.]{
\centering
\includegraphics[width=0.4\columnwidth]{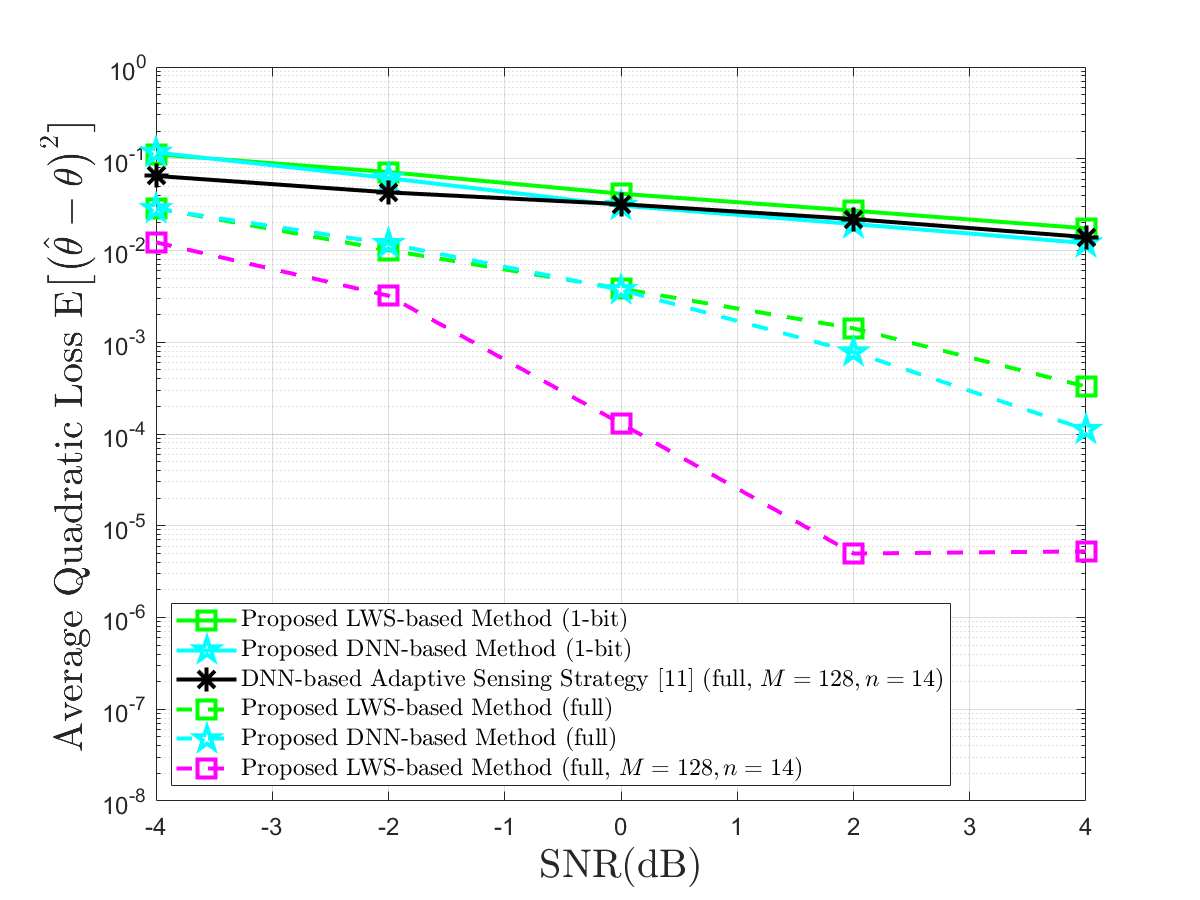}
\label{quadratic_loss_full}
}
\caption{Performance comparison between the benchmark algorithms and the two proposed methods. As observed, our proposed methods under the 1-bit measurement rule are comparable to the hierarchical beam sweeping algorithm, and far better than the naive beam sweeping algorithm, while our proposed methods under the full measurement rule outperform all benchmark algorithms. Furthermore, for both measurement rules, the DNN-based method slightly outperforms the LWS-based method.}
\label{quadratic_loss}
\end{figure}

In Fig. \ref{quadratic_loss_1bit}, we plot the average quadratic loss $\mathbb{E}[(\hat{\theta}-\theta)^2]$ with 1-bit measurement rule for the two proposed methods, computed over $10^3$ Monte Carlo runs. Due to the phenomenon of flattening the entire power spectrum in the DNN-based method mentioned in Section \ref{mr Analysis of Two Methods with 1-bit Measurement Rule}, we only use it to train questioners when $K \leq \frac{M}{4}$, which is $16$ in this example. It can be seen that the performance of the LWS-based method (green curve with squares) is comparable to the hierarchical beam sweeping algorithm (blue curve with triangles), and far better than the naive beam sweeping algorithm (red curve with circles). Furthermore, the performance of the DNN-based method (cyan curve with pentagrams) can outperform (by at most $1.5$dB in this example) the LWS-based method, and continuously improves as the SNR increases, which matches the results in Fig. \ref{lws_vs_dnn_1bit} that the DNN-based method has the potential to achieve better performance, especially when the SNR is high. 

Here we provide another perspective on why the outperformance occurs. Recall that both methods are different implementations of the sortPM algorithm~\cite{chiu2021low} under query-dependent channels for the adaptive BA in practical communication systems. A major characteristic of the sortPM algorithm is that it can achieve optimal performance when the noise is positively correlated with the query size, that is, when the noise increases as the query size grows. On the contrary, when the noise is negatively correlated with the query, the algorithm performance will significantly deteriorate. Recall the simulation in Fig. \ref{1bit_power}. For the DNN-based method, as the number of sub-intervals within the search region increases, the gap widens, and the probability of mismatches due to imperfectly trained questioners also increases. While for the LWS-based method, the gap narrows as the number of sub-intervals $K$ within the search region increases (when $K \geq 3$), resulting in lower performance compared to the DNN-based method.

We now plot the average quadratic loss $\mathbb{E}[(\hat{\theta}-\theta)^2]$ with full measurement rule for the two proposed methods, also computed over $10^3$ Monte Carlo runs. In order to maintain consistency with the simulation in Fig. \ref{quadratic_loss_1bit}, the DNN-based method is also used when $K \leq \frac{M}{4}$. Fig. \ref{quadratic_loss_full} shows that when the channel condition is poor, the two proposed methods show little difference; when the channel condition is good, the DNN-based method (cyan dashed curve with pentagrams) slightly outperforms (by at most $0.8$dB in this example) the LWS-based method (green dashed curve with squares). Furthermore, to maintain consistency with the simulation settings in~\cite{sohrabi2021deep}, we also plot for the LWS-based method with the maximum number of samples $n = 14$ and the number of sub-intervals $M = 128$, which shows that the LWS-based method (magenta dashed curve with squares) is far better than the DNN-based adaptive sensing strategy (black curve with star) with full measurement rule. When $\mathrm{SNR} \geq 2$, the reason for the plateau of the magenta dashed curve with squares is that the average quadratic loss has reached the performance limit under the setting of $M=128$.

Furthermore, we can notice that the performance with full measurement rule is also far better than that with 1-bit measurement rule, which can be explained by comparing the power spectrum in Fig. \ref{power_spec_lws_overall} and the spectrum of the probability factor in Fig. \ref{probFactor_spec}. Compared with the power spectrum, the three peaks of the spectrum of the probability factor are narrower in width, which makes each update of the posterior probability vector more rapid and accurate, thereby resulting in a smaller quadratic loss of AoA estimation in the adaptive BA algorithm.

\subsection{The Case of Unknown Fading Coefficient $\alpha$} \label{nr The Case of Unknown Fading Coefficient}

The above simulations are running with the assumption that the fading coefficient $\alpha$ is available to RX. However, a more practical scenario is when $\alpha$ is unknown. For the unknown fading coefficient case, there are two estimators proposed in~\cite{sohrabi2021deep}, namely minimum mean squared error (MMSE) estimator and Kalman filter estimator, respectively. It has been demonstrated that the Kalman filter estimator is computationally more efficient than the MMSE estimator, offering storage and computational advantages. This inspires us to investigate whether this approach can be adopted here.

Let $\bc_{i,t}=\bw_t^H \bA(\theta_i)$. Assume that the unknown fading coefficient $\alpha$ follows a complex Gaussian distribution with mean $\mu_{\alpha}$ and variance $\sigma_{\alpha}$. Recall the definition of the received signal $\by_t$ in \eqref{r-signal} at any time point $t \in \bbN$. For any presumed AoA $\theta_i \in [\theta_{\text{min}},\theta_{\text{max}}]$, the corresponding mean $\mu_{i,t}^{\alpha}$ and variance $\sigma_{i,t}^{\alpha}$ of $\alpha$ can be updated by Kalman filter~\cite{kay1993fundamentals}:
\begin{align}
\mu_{i,t+1}^{\alpha}&=\mu_{i,t}^{\alpha}+\frac{\sigma_{i,t}^{\alpha}\bc_{i,t}^*}{\sigma_{i,t}^{\alpha}|\bc_{i,t}|^2+1}\big(\by_t-\mu_{i,t}^{\alpha}\bc_{i,t}\big),\\
\sigma_{i,t+1}^{\alpha}&=\frac{\sigma_{i,t}^{\alpha}}{\sigma_{i,t}^{\alpha}|\bc_{i,t}|^2+1},
\end{align}
where $\bc_{i,t}^*$ denotes the conjugate of $\bc_{i,t}$.

Therefore, under the assumption of unknown fading coefficient, the probability factor $\nu$ defined in \eqref{nu} will be replaced as follows:
\begin{align}
    \nu(\theta_i,\by_t,\bw_t) := e^{-\frac{|\by_t-\mu_{i,t}^{\alpha}\bc_{i,t}|^2}{\sigma_{i,t}^{\alpha}|\bc_{i,t}|^2+1}}. \label{nu-unknown}
\end{align}

\begin{figure}[tb]
\centering
\subfloat[Plot of average quadratic loss versus SNR for the LWS-based method.]{
\centering
\includegraphics[width=0.4\textwidth]{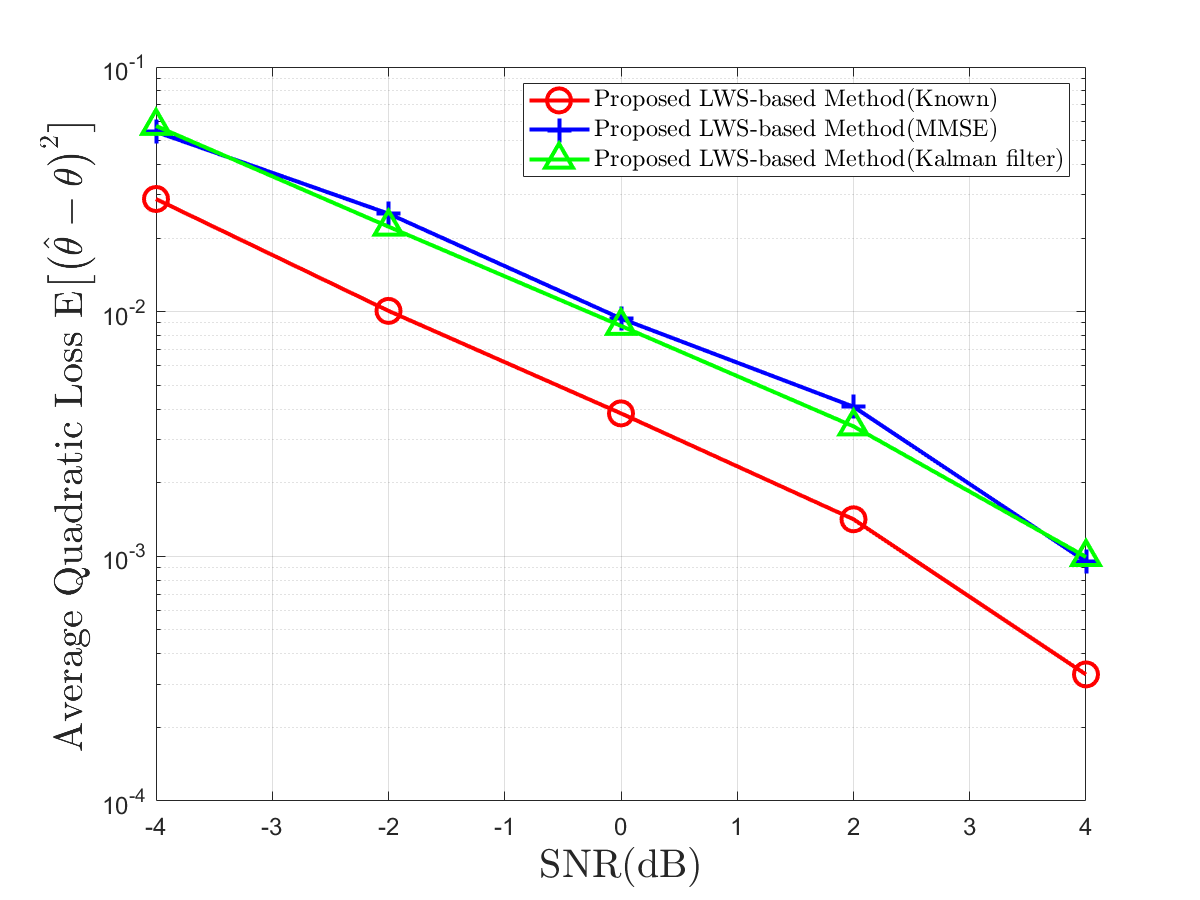}
\label{unknown_lws}
}
\hfill
\subfloat[Plot of average quadratic loss versus SNR for the DNN-based method.]{
\centering
\includegraphics[width=0.4\columnwidth]{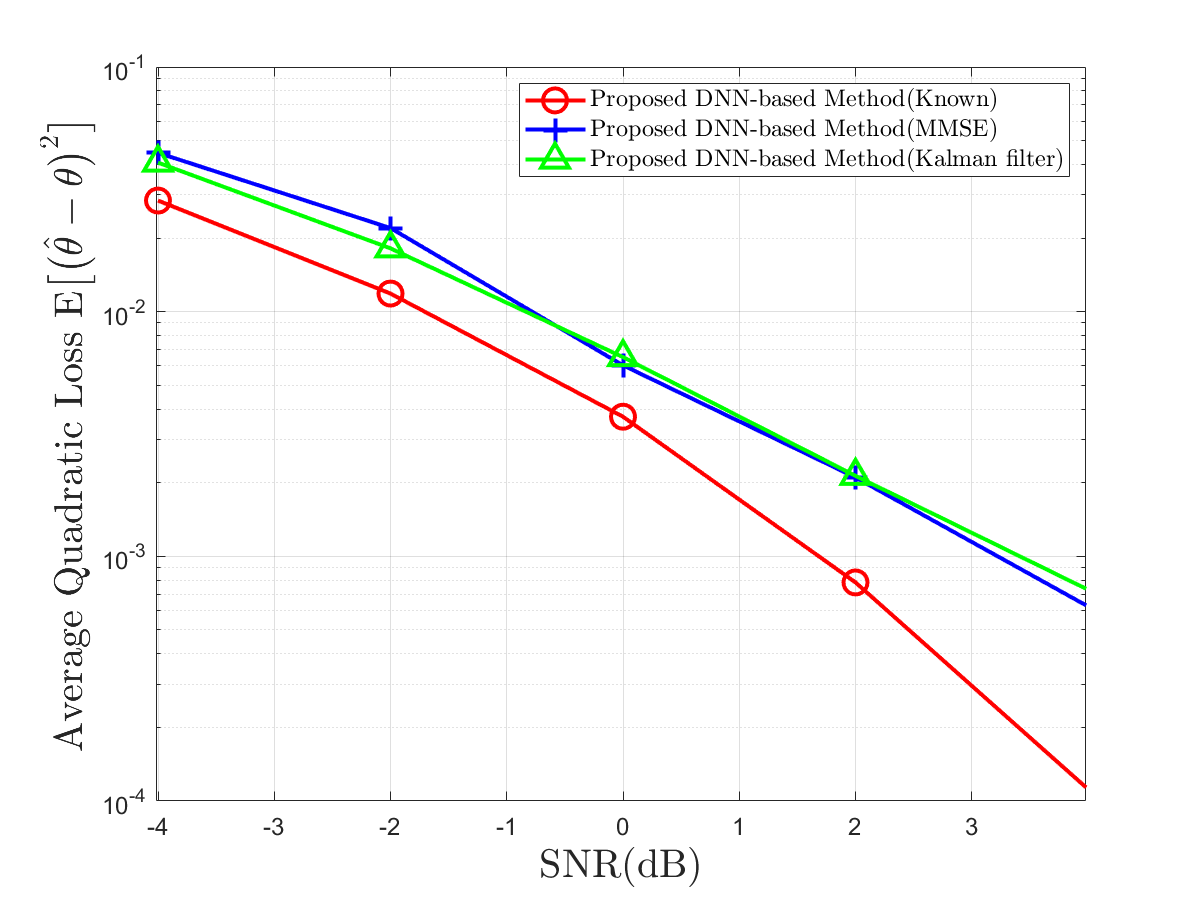}
\label{unknown_dnn}
}
\caption{Performance comparison for two proposed methods in the three cases: known $\alpha$, unknown $\alpha$ with MMSE estimator and unknown $\alpha$ with Kalman filter estimator. As observed, when the fading coefficient $\alpha$ is unknown, two proposed method can still achieve good performance with the help of both MMSE and Kalman filter estimators, with a performance loss of only up to 2dB. Furthermore, the performance loss caused by the two estimators is comparable.}
\label{quadratic_loss_unknown}
\end{figure}

\begin{figure}[tb]
\centering
\includegraphics[width=0.6\columnwidth]{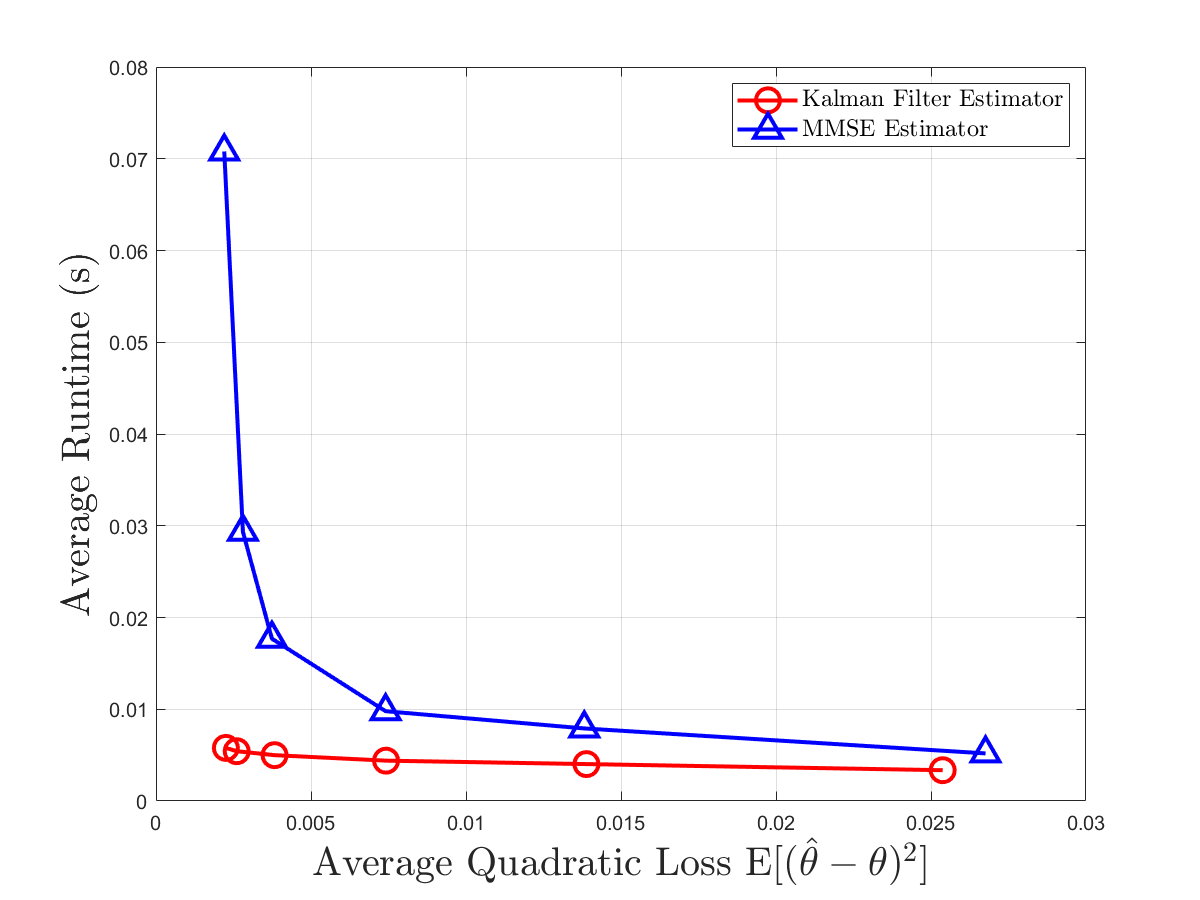}
\caption{Plot of average runtime versus average quadratic loss $\mathbb{E}[(\hat{\theta}-\theta)^2]$ for MMSE estimator and Kalman filter estimator with the LWS-based method. As observed, the runtime of the Kalman filter estimator is significantly less than the MMSE estimator, and the gap between the two significantly increases with the decrease of $\mathbb{E}[(\hat{\theta}-\theta)^2]$.}
\label{unknown_runtime}
\end{figure}

To demonstrate the feasibility of the Kalman filter estimator on the two proposed methods, we also plot the average quadratic loss $\mathbb{E}[(\hat{\theta}-\theta)^2]$ with full measurement rule under the setting of $\alpha \sim \mathcal{CN}(1,0.25^2)$ over $10^3$ Monte Carlo runs. For the consistency, we only replace the definition of the probability factor $\nu$ in \eqref{nu} with \eqref{nu-unknown}. As shown in Fig. \ref{quadratic_loss_unknown}, both estimator are applicable to with acceptable performance loss. Furthermore, as a measure of storage and computational complexity, Fig. \ref{unknown_runtime} shows that the average runtime of the Kalman filter estimator is significantly less than the MMSE estimator for the same average quadratic loss, which is consistent with the analysis of the advantages of Kalman filter estimator in~\cite{sohrabi2021deep}.

\section{Conclusion} \label{sec_conc}
We proposed adaptive BA algorithms using the noisy twenty questions estimation framework with trained questioners, addressing the infeasibility of the algorithm in \cite{chiu2019active} by specifying how to convert the query region of twenty questions into the corresponding beamforming vector for BA. In particular, we proposed a method based on LWS and another method of training NNs. Via numerical simulations, we analyzed the reasons for the performance gaps between the questioners trained by the two proposed methods and the questioner in the ideal system model in \cite{chiu2019active}. Furthermore, we explained the rationality of query-dependent channels in the system model, discussed the reasons for the performance gap between 1-bit measurement rule and full measurement rule and showed that our proposed BA algorithms significantly outperforms the existing adaptive BA algorithms.

There are several other avenues for future directions. Firstly, in this paper, we proposed two methods for adaptive BA, aiming to simultaneously achieve feasibility, interpretability and low latency. However, there remains a gap between their performance and the theoretical optimum. One can explore other non-NN or NN methods to narrow this gap, potentially even reaching the theoretical limits. Secondly, we focused on the LoS setting in this paper, i.e., there is only one AoA to be estimated. But non-LoS setting is more common in practice, involving multiple AoAs to be estimated. In that case, the optimal beamforming vector should exhibit a multi-cluster structure~\cite{raghavan2016beamforming}. It is thus of interest to extend the active learning framework to the multi-target case, derive the corresponding theoretical results and apply it to the non-LoS setting. Thirdly, we only considered the sortPM strategy proposed in~\cite[Eq. (13)]{chiu2021low} in this paper, while introducing alternative strategies, such as hiePM and dyaPM, may yield new insights and better performance. It would also be interesting to investigate the cases where sortPM is replaced with hiePM and dyaPM in our proposed algorithms. Furthermore, we followed the same strategy as in~\cite[Eq. (13)]{chiu2021low} to generate the queries. It may lead to a mismatch between the actual distribution of the generated codebook and the capacity-achieving input distribution, which can degrade the algorithms' performance. It is worthwhile to find out how to adjust the strategy to eliminate the mismatch, or how to quantify the performance loss if the mismatch is unavoidable. Finally, we focused on the stationary target case in this paper. However, a target in motion represents a more realistic setting in communication systems. Our results of the adaptive BA for a stationary target can be generalized to the moving target case via ideas in~\cite{ronquillo2023integrated,zhou2023resolution,zhou2025twenty,sun2025resolutionlimitsnonadaptive20}.

\bibliographystyle{IEEEtran}
\bibliography{IEEEabrv,IEEE_cs}

\end{document}

%% file: preamble.tex
\usepackage{soul}
\usepackage[mathscr]{eucal}
\usepackage[cmex10]{amsmath}
\usepackage{epsfig,epsf,psfrag}
\usepackage{amssymb,amsmath,amsthm,amsfonts,latexsym}
\usepackage{amsmath,graphicx,bm,xcolor,url,overpic}
\usepackage[caption=false]{subfig} 
\usepackage{fixltx2e}
\usepackage{array}
\usepackage{verbatim}
\usepackage{bm}
\usepackage{algorithmic}
\usepackage{algorithm}
\usepackage{verbatim}
\usepackage{textcomp}
\usepackage{mathrsfs}
\usepackage{epstopdf}
\usepackage{setspace}

\newcommand{\openone}{\leavevmode\hbox{\small1\normalsize\kern-.33em1}}

\catcode`~=11 \def\UrlSpecials{\do\~{\kern -.15em\lower .7ex\hbox{~}\kern .04em}} \catcode`~=13 

\allowdisplaybreaks[4]

\newcommand{\calA}{\mathcal{A}}

\newcommand{\calC}{\mathcal{C}}
\newcommand{\calD}{\mathcal{D}}

\newcommand{\calF}{\mathcal{F}}

\newcommand{\calI}{\mathcal{I}}

\newcommand{\calP}{\mathcal{P}}

\newcommand{\calS}{\mathcal{S}}

\newcommand{\calX}{\mathcal{X}}
\newcommand{\calY}{\mathcal{Y}}

\newcommand{\ba}{\mathbf{a}}
\newcommand{\bA}{\mathbf{A}}

\newcommand{\bc}{\mathbf{c}}

\newcommand{\bG}{\mathbf{G}}
\newcommand{\bh}{\mathbf{h}}

\newcommand{\bI}{\mathbf{I}}

\newcommand{\bn}{\mathbf{n}}

\newcommand{\bS}{\mathbf{S}}

\newcommand{\bw}{\mathbf{w}}

\newcommand{\by}{\mathbf{y}}

\newcommand{\rma}{\mathrm{a}}

\newcommand{\rmb}{\mathrm{b}}

\newcommand{\rmm}{\mathrm{m}}

\newcommand{\rmq}{\mathrm{q}}

\newcommand{\rmu}{\mathrm{u}}

\newcommand{\bbC}{\mathbb{C}}

\newcommand{\bbN}{\mathbb{N}}

\newcommand{\bbP}{\mathbb{P}}

\newcommand{\bbR}{\mathbb{R}}

\newcommand{\frakA}{\mathfrak{A}}
\newcommand{\frakB}{\mathfrak{B}}

\newcommand{\frakI}{\mathfrak{I}}

\newcommand{\frakR}{\mathfrak{R}}

\DeclareMathAlphabet{\mathbsf}{OT1}{cmss}{bx}{n}
\DeclareMathAlphabet{\mathssf}{OT1}{cmss}{m}{sl}

\DeclareSymbolFont{bsfletters}{OT1}{cmss}{bx}{n}  
\DeclareSymbolFont{ssfletters}{OT1}{cmss}{m}{n}
\DeclareMathSymbol{\bsfGamma}{0}{bsfletters}{'000}
\DeclareMathSymbol{\ssfGamma}{0}{ssfletters}{'000}
\DeclareMathSymbol{\bsfDelta}{0}{bsfletters}{'001}
\DeclareMathSymbol{\ssfDelta}{0}{ssfletters}{'001}
\DeclareMathSymbol{\bsfTheta}{0}{bsfletters}{'002}
\DeclareMathSymbol{\ssfTheta}{0}{ssfletters}{'002}
\DeclareMathSymbol{\bsfLambda}{0}{bsfletters}{'003}
\DeclareMathSymbol{\ssfLambda}{0}{ssfletters}{'003}
\DeclareMathSymbol{\bsfXi}{0}{bsfletters}{'004}
\DeclareMathSymbol{\ssfXi}{0}{ssfletters}{'004}
\DeclareMathSymbol{\bsfPi}{0}{bsfletters}{'005}
\DeclareMathSymbol{\ssfPi}{0}{ssfletters}{'005}
\DeclareMathSymbol{\bsfSigma}{0}{bsfletters}{'006}
\DeclareMathSymbol{\ssfSigma}{0}{ssfletters}{'006}
\DeclareMathSymbol{\bsfUpsilon}{0}{bsfletters}{'007}
\DeclareMathSymbol{\ssfUpsilon}{0}{ssfletters}{'007}
\DeclareMathSymbol{\bsfPhi}{0}{bsfletters}{'010}
\DeclareMathSymbol{\ssfPhi}{0}{ssfletters}{'010}
\DeclareMathSymbol{\bsfPsi}{0}{bsfletters}{'011}
\DeclareMathSymbol{\ssfPsi}{0}{ssfletters}{'011}
\DeclareMathSymbol{\bsfOmega}{0}{bsfletters}{'012}
\DeclareMathSymbol{\ssfOmega}{0}{ssfletters}{'012}

\DeclareMathOperator*{\argmax}{arg\,max}

\newtheorem{definition}{Definition}